\documentclass[fleqn,usenatbib]{mnras}

\usepackage{newtxtext,newtxmath}
\usepackage{fontawesome5}

\usepackage[T1]{fontenc}

\DeclareRobustCommand{\VAN}[3]{#2}
\let\VANthebibliography\thebibliography
\def\thebibliography{\DeclareRobustCommand{\VAN}[3]{##3}\VANthebibliography}

\usepackage{graphicx}	

\usepackage[normalem]{ulem}
\usepackage{adjustbox} 

\usepackage{color}
\usepackage{acronym}   
\usepackage{xspace}
\usepackage{calc}
\usepackage[caption=false]{subfig}
\usepackage{wrapfig} 
\usepackage{import} 
\usepackage{multirow}
\usepackage{caption}

\definecolor{tomcol}{rgb}{0.53,0.00,1.00}

\definecolor{ochre}{rgb}{0.8, 0.47, 0.13}

\usepackage{outlines} 

\newcommand\rate{\mathcal{R}}
\newcommand\COMPAS{{\sc{COMPAS }}}

\newcommand{\monei}{\ensuremath{m_{1,\rm{i}}}\xspace}

\newcommand{\Zi}{\ensuremath{Z}\xspace}
\newcommand{\vk}{\ensuremath{v_{\rm{k}}}\xspace}

\newcommand{\sigmacc}{\ensuremath{\sigma_{\rm{rms}}^{\rm{1D}}\xspace}}

\newcommand{\kms}{\ensuremath{\,\rm{km}\,\rm{s}^{-1}}\xspace}
\newcommand{\Msun}{\ensuremath{\,\rm{M}_{\odot}}\xspace}

\newcommand{\Zsun}{\ensuremath{\,\rm{Z}_{\odot}}\xspace}

\newcommand{\Gyr}{\ensuremath{\,\mathrm{Gyr}}\xspace}

\newcommand{\yearmin}{\ensuremath{\,\rm{yr}^{-1}}\xspace}

\newcommand{\GpcminThree}{\ensuremath{\,\rm{Gpc}^{-3}}\xspace}

\newcommand{\MSFR}{\ensuremath{{M}_{\rm{SFR}}}\xspace}
\newcommand{\SFRD}{\ensuremath{\mathcal{S}(Z,z)}\xspace}
\newcommand{\tdelay}{\ensuremath{{t}_{\rm{delay}}}\xspace}

\newcommand{\ts}{\ensuremath{{t}_{\rm{s}}}\xspace}
\newcommand{\tevolve}{\ensuremath{{t}_{\rm{evolve}}}\xspace}
\newcommand{\tform}{\ensuremath{{t}_{\rm{form}}}\xspace}
\newcommand{\tmerger}{\ensuremath{{t}_{\rm{m}}}\xspace}
\newcommand{\tinspiral}{\ensuremath{{t}_{\rm{inspiral}}}\xspace}

\newcommand{\tdet}{\ensuremath{{t}_{\rm{det}}}\xspace}
\newcommand{\Nform}{\ensuremath{{N}_{\rm{form}}}\xspace}
\newcommand{\Ndet}{\ensuremath{{N}_{\rm{det}}}\xspace}
\newcommand{\Nmerger}{\ensuremath{{N}_{\rm{merger}}}\xspace}

\newcommand{\Pdet}{\ensuremath{{P}_{\rm{det}}}\xspace}

\newcommand{\Vc}{\ensuremath{{V}_{\rm{c}}}\xspace}

\newcommand*\diff{\mathop{}\!\mathrm{d}}
\newcommand{\mnsf}{\ensuremath{m_{\rm{NS}}}\xspace}
\newcommand{\mnsfone}{\ensuremath{m_{\rm{NS,1}}}\xspace}
\newcommand{\mnsftwo}{\ensuremath{m_{\rm{NS,2}}}\xspace}
\newcommand{\mbhf}{\ensuremath{m_{\rm{BH}}}\xspace}
\newcommand{\mbhfone}{\ensuremath{m_{\rm{BH,1}}}\xspace}

\newcommand{\mchirpf}{\ensuremath{{\mathcal{M}}_{\rm{c}}}\xspace}

\newcommand{\qf}{\ensuremath{q}\xspace}

\newcommand{\Nmodels}{\ensuremath{560}\xspace}
\newcommand{\NmodelsBPS}{\ensuremath{20}\xspace}
\newcommand{\NmodelsMSSFR}{\ensuremath{28}\xspace}

\newcommand{\RateIntrinsicZero}{\ensuremath{\mathcal{R}_{\rm{m}}^{0}}\xspace}
\newcommand{\RateObserved}{\ensuremath{\mathcal{R}_{\rm{det}}}\xspace}

\newcommand{\RateIntrinsicAzeroBHBHmin}{\ensuremath{3.8}\xspace} 
\newcommand{\RateIntrinsicAzeroBHBHmax}{\ensuremath{810}\xspace} 
\newcommand{\RateIntrinsicAzeroBHNSmin}{\ensuremath{2.2}\xspace}  
\newcommand{\RateIntrinsicAzeroBHNSmax}{\ensuremath{830}\xspace} 
\newcommand{\RateIntrinsicAzeroNSNSmin}{\ensuremath{0.32}\xspace} 
\newcommand{\RateIntrinsicAzeroNSNSmax}{\ensuremath{330}\xspace} 

\newcommand{\RateObservedAzeroBHBHmax}{\ensuremath{12000}\xspace} 
\newcommand{\RateObservedAzeroBHBHmin}{\ensuremath{17}\xspace} 
\newcommand{\RateObservedAzeroBHNSmax}{\ensuremath{180}\xspace} 
\newcommand{\RateObservedAzeroBHNSmin}{\ensuremath{0.36}\xspace} 
\newcommand{\RateObservedAzeroNSNSmax}{\ensuremath{12}\xspace} 
\newcommand{\RateObservedAzeroNSNSmin}{\ensuremath{0}\xspace}

\newcommand{\RateGWTCmaxBHBH}{\ensuremath{130}\xspace} 
\newcommand{\RateGWTCminBHBH}{\ensuremath{16}\xspace} 
\newcommand{\RateGWTCmaxBHNS}{\ensuremath{320}\xspace} 
\newcommand{\RateGWTCminBHNS}{\ensuremath{7.4}\xspace} 
\newcommand{\RateGWTCmaxNSNS}{\ensuremath{1900}\xspace} 
\newcommand{\RateGWTCminNSNS}{\ensuremath{16}\xspace}

\acrodef{GSMF}{galaxy stellar mass function, the number density of galaxies per logarithmic mass bin,}
\acrodef{MZR}{mass-metallicity relation}
\acrodef{SFRD}{star formation rate density}
\acrodef{BHNS}{black hole--neutron star}
\acrodef{NSNS}{binary neutron star}
\acrodef{BHBH}{binary black hole}
\acrodef{DCO}{double compact object}
\acrodef{NS}{neutron star}
\acrodef{BH}{black hole}
\acrodef{BH--NS}{black hole-neutron star}
\acrodef{GRB}{gamma-ray burst}
\acrodef{RLOF}{Roche-lobe overflow}
\acrodef{CE}{common envelope}
\acrodef{GW}{gravitational-wave}
\acrodefplural{GW}[GWs]{gravitational waves}
\acrodef{SN}{supernova}
\acrodefplural{SN}[SNe]{supernovae}
\acrodef{ECSN}{electron-capture SN}
\acrodef{PISN}{pair-instability SN}
\acrodefplural{ECSN}[ECSNe]{electron-capture SN}
\acrodef{USSN}{ultra-stripped SN}
\acrodefplural{USSN}[USSNe]{ultra-stripped SN}
\acrodef{CCSN}{core-collapse SN}
\acrodefplural{CCSN}[CCSNe]{core-collapse SN}
\acrodef{COMPAS}{
Compact Object Mergers: Population Astrophysics and Statistics}
\defcitealias{Broekgaarden:2021}{Paper I}

\title[Gravitational Wave Rates and Distribution Shapes]{Impact of Massive Binary Star and Cosmic Evolution on Gravitational Wave Observations II: Double Compact Object Rates and Properties}

\author[Broekgaarden et al.]{{Floor S. Broekgaarden},$^{1}$\thanks{E-mail: floor.broekgaarden@cfa.harvard.edu}
{Edo Berger},$^{1}$
Simon Stevenson,$^{2,3}$
{Stephen Justham},$^{4,5,6,7}$
{Ilya Mandel},$^{8,3,9}$
\newauthor
{Martyna Chru{\'s}li{\'n}ska},$^{7,10}$
Lieke A. C. van Son,$^{1,6,7}$
Tom Wagg,$^{11,1,7}$ 
{Alejandro Vigna-G\'{o}mez},$^{12,13}$
\newauthor
{Selma E. de Mink},$^{7,6,1}$
Debatri Chattopadhyay,$^{2,3}$
{Coenraad J. Neijssel}$^{9}$
\\
$^{1}${Center for Astrophysics \textbar{} Harvard $\&$ Smithsonian,
60 Garden St., Cambridge, MA 02138, USA}\\
$^{2}${Center for Astrophysics and Supercomputing, Swinburne University of Technology, Hawthorn VIC 3122, Australia}\\
$^{3}${The ARC Center of Excellence for Gravitational Wave Discovery -- OzGrav, Hawthorn VIC 3122, Australia}\\
$^{4}${School of Astronomy $\&$ Space Science, University of the Chinese Academy of Sciences, Beijing 100012, China}\\
$^{5}${National Astronomical Observatories, Chinese Academy of Sciences, Beijing 100012, China} \\
$^{6}${Anton Pannekoek Institute for Astronomy and GRAPPA, University of Amsterdam, Postbus 94249, 1090 GE Amsterdam, The Netherlands }\\
$^{7}${Max-Planck-Institut f\"{u}r Astrophysik, Karl-Schwarzschild-Stra{\ss}e 1, 85741 Garching, Germany} \\
$^{8}${Monash Centre for Astrophysics, School of Physics and Astronomy, Monash University, Clayton, Victoria 3800, Australia}\\
$^{9}${Birmingham Institute for Gravitational Wave Astronomy and School of Physics and Astronomy, University of Birmingham, Birmingham, B15 2TT, United Kingdom}\\
$^{10}${Institute of Mathematics, Astrophysics and Particle Physics, Radboud University Nijmegen, PO Box 9010, 6500 GL Nijmegen}\\
$^{11}${Department of Astronomy, University of Washington, Seattle, WA, 98195}\\
$^{12}${Niels Bohr International Academy, The Niels Bohr Institute, Blegdamsvej 17, DK-2100 Copenhagen, Denmark}\\
$^{13}${DARK, Niels Bohr Institute, University of Copenhagen, Jagtvej 128, DK-2200, Copenhagen, Denmark}}

\date{Accepted XXX. Received YYY; in original form ZZZ}

\pubyear{2021}

\begin{document}
\label{firstpage}
\pagerange{\pageref{firstpage}--\pageref{lastpage}}
\maketitle

\begin{abstract}
Making the most of the rapidly increasing population of gravitational-wave detections of \ac{BH} and \ac{NS} mergers requires comparing observations with population synthesis predictions.  In this work we investigate the \textit{combined} impact from the key uncertainties in population synthesis modelling of the isolated binary evolution channel: the physical processes in massive binary-star evolution and the star formation history as a function of metallicity, $Z$, and redshift $z$, \SFRD. Considering these uncertainties we create \Nmodels different publicly available model realizations and calculate the rate and distribution characteristics of detectable BHBH, BHNS, and NSNS mergers. 
We find that our stellar evolution and \SFRD variations can impact the predicted intrinsic and detectable merger rates by factors $10^2$--$10^4$. We find that {BHBH} rates are dominantly impacted by \SFRD variations, {NSNS} rates by stellar evolution variations and {BHNS} rates by both.
We then consider the combined impact from all uncertainties considered in this work on the detectable mass {distribution shapes}  (chirp mass, individual masses and mass ratio). We find that the {BHNS} mass distributions are predominantly impacted by massive binary-star evolution changes. For {BHBH} and {NSNS} we find that both uncertainties are important. 
We also find that the shape of the delay time and birth metallicity distributions are typically dominated by the choice of \SFRD for BHBH, BHNS and NSNS. 
We identify several examples of robust features in the mass distributions predicted by all \Nmodels models, such that we expect more than 95 percent of {BHBH} detections to contain a \ac{BH} $\gtrsim 8\Msun$ and have mass ratios $ \lesssim 4$.  
Our work demonstrates that it is essential to consider a wide range of allowed models to study double compact object merger rates and properties.  Conversely, larger observed samples could allow us to decipher currently unconstrained stages of stellar and binary evolution.

\end{abstract}





\section{Introduction}
\label{sec:introduction}

The population of detected \ac{GW} events from \ac{BHBH}, \ac{BHNS} and \ac{NSNS} mergers is rapidly increasing \citep{2019PhRvX...9c1040A,2020LRR....21....3A,Abbott:GWTC-2,Abbott:2021GWTC3}.  
These mergers carry unique information about the properties of \acp{BH} and \acp{NS} (such as their masses and spins), which in turn probes the formation, lives, and explosive deaths of massive stars throughout cosmic history \citep[e.g.][]{Abbott:2021GWTC2pop,Abbott:2021GWTC3pop}. 
To extract information  from these  \ac{DCO} detections requires comparing their observed properties, to theoretically simulated populations modelling their formation pathways.

A variety of formation channels for \ac{DCO} mergers have been proposed \citep[see][for  reviews]{MandelFarmer:2018,Mapelli:2021review}, with 
one of the most prominent pathways being the \emph{isolated binary evolution channel}, where the merging \ac{DCO} systems are assumed to form from pairs of massive stars in wide, isolated, binary systems. 
This formation pathway can currently account for the observed \ac{DCO} rates \citep[][]{MandelBroekgaardenReview:2021} and many of the observed \ac{DCO} source properties (e.g., \citealt[][]{VignaGomez:2018,Belczynski:2020bigBHpaper}, \citealt{BroekgaardenBerger2021}, but see \S\ref{sec:discussion-comparison-with-GW-observations} for a discussion).

However, modelling theoretical populations of \ac{DCO} mergers from the isolated binary evolution channel is challenging as the simulations suffer from two key  uncertainties:  First, various physical processes in massive binary star evolution are uncertain; these include key evolutionary stages such as wind mass loss, stable mass transfer, \ac{CE} episodes, and \acp{SN}, which significantly impact the predicted rates and properties of \ac{DCO} mergers.  
Second, there are critical uncertainties in the cosmic star formation and chemical evolution history, which impact the metallicity-dependent star formation rate density \SFRD, which is a function of birth metallicity \Zi and redshift $z$. 
Uncertainties in \SFRD also significantly impact the predicted rate and properties of \ac{DCO} mergers as the birth metallicity significantly affects stellar evolution, including mass loss through line-driven winds and the (maximum) radial expansion.

Earlier works investigating the impact from stellar evolution and \SFRD uncertainties on the simulated \ac{DCO} population typically focused on exploring only one of these two sets of uncertainties.  
For example, studies including \citet{Dominik:2015}, \citet{GiacobboMapelli:2018}, \citet{Kruckow:2018}, and \citet{Belczynski:2020bigBHpaper} explored population synthesis models with a large number of different massive binary stellar evolution assumptions, but only investigated one or a few \SFRD models. 
On the other hand, studies including \citet{Chruslinska:2019dco}, \citet{Neijssel:2019}, \citet{Tang:2020}, and \citet{Briel:2021} focused on exploring a wide range of \SFRD models, but only considered one or a few different stellar evolution realizations. 
By only focusing on one of the two uncertainties it remains challenging to directly understand the combined or relative impact from the massive (binary) stellar evolution and \SFRD uncertainties on the detectable \ac{DCO} population. 
This limits our ability to learn from \ac{GW} observations and constrain the models.

Recently, \citet{Santoliquido:2021} improved on this by presenting a large set of  binary star evolution model variations (for mass transfer, \ac{CE} phases and \acp{SN}) as well as \SFRD models, exploring the impact from each of the uncertainties on the predicted \ac{DCO} merger \emph{rates} (although the authors do not present the \emph{combined} impact from both uncertainties except for their $\alpha1$ and  $\alpha5$ model variations in their Figure~7). 
In addition, \citet{Chu:2021} carried out a large study investigating the impact from variations in the \ac{CE} phase and \ac{SN} kicks in combination with four different \SFRD models focusing on their impact on the merger rates of \ac{NSNS} mergers. 
However, besides impacting the merger rates, variations in stellar evolution models and \SFRD are also expected to impact the \ac{DCO} merger \emph{distribution shapes} of the properties of the detectable mergers. 
A large systematic study exploring this for \ac{GW} events from all three types of \ac{DCO} mergers (\ac{BHBH}, \ac{BHNS} and \ac{NSNS}) is currently missing.

In \citet[][from hereon  \citetalias{Broekgaarden:2021}]{Broekgaarden:2021},  we improved on this by investigating the \emph{combined} impact from uncertainties in both massive binary star evolution and the \SFRD on the predicted \emph{rate} and \emph{distribution shapes} focusing on, as a first step, the \ac{GW}-detectable \ac{BHNS} mergers. Here we continue this work by studying all three \ac{GW}-detectable \ac{DCO} merger types and by adding extra model realizations.
We use rapid binary stellar evolution synthesis simulations coupled with analytical prescriptions for \SFRD to present the \ac{DCO} rate and properties for a total of \Nmodels model realizations. Our method is described in \S\ref{sec:method}. We present the impact from the physical processes in massive binary star evolution and \SFRD on the predicted rate and shape of the distribution functions for \ac{DCO} mergers in \S\ref{section:results}. We discuss these results in \S\ref{sec:discussion} and present our conclusions in \S\ref{sec:conclusions}. 


\section{Method}
\label{sec:method}
We use the simulations and methodology presented in  \citetalias{Broekgaarden:2021}. Here we add new models F, G, J, S and T, which leads to a minor shift in the model labels. Our models explore the key assumptions in both massive (binary) star evolution and \SFRD, leading to \Nmodels  (\NmodelsBPS binary stellar evolution $\times$ \NmodelsMSSFR \SFRD) model realizations. We particularly choose our model variations to explore a broad span of uncertainty in the modelling. For that reason some of the variations are somewhat extreme (e.g., models S and T), but are chosen to explore their possible impact on the rate and distribution shapes of \ac{DCO} mergers. We summarize our most important model assumptions below and summarize the model variations in Table~\ref{tab:variations-BPS}  (massive binary star evolution) and Table~\ref{tab:MSSFR-variations-labels} (\SFRD).  More details can be found in the Appendix and in  \citetalias{Broekgaarden:2021}.

\subsection{Massive binary-star population models}
\label{sec:method-binary-population-synthesis-model-assumptions}

We simulate populations of \ac{GW} sources with the rapid binary population synthesis code from the~{\sc{COMPAS}}\footnote{Compact Object Mergers: Population Astrophysics and Statistics,  \url{https://compas.science}} suite \citep{stevenson2017formation, 2018MNRAS.477.4685B, VignaGomez:2018, Broekgaarden:2019, Neijssel:2019,COMPAS:2021methodsPaper}. We model the formation of \ac{BH} and \ac{NS} mergers from the \textit{isolated binary evolution} channel where the merging \acp{DCO} form from massive stars born in a binary system \citep{1976ApJ...207..574S, 1989A&ARv...1..209S}. We explore uncertainties in our massive (binary) star evolution assumptions by studying \NmodelsBPS binary population synthesis model variations.  We further use the efficient sampling algorithm STROOPWAFEL \citep{Broekgaarden:2019} to simulate $\gtrsim 10^6$ binary systems for 53 different \Zi values, resulting in typically $\sim 10^7$ \ac{DCO} mergers in each simulation, making it one of the best-sampled population synthesis studies of its kind. 

For our default model (A) settings we use the fiducial model summarized in  \citetalias{Broekgaarden:2021}.
For the remaining models we change one population parameter at a time (relative to the fiducial model) to explore the impact of the uncertain model assumptions. The only exception is model F (E+K), which combines the model changes from both models E and K. Our \NmodelsBPS models are summarized in Table~\ref{tab:variations-BPS} and we refer to them throughout the rest of the paper by the letters A to T and the abbreviated label names given in the second column of Table~\ref{tab:variations-BPS}.  We focus on (and decide to use) the stellar evolution and the \SFRD variations mentioned below as these are commonly used in population synthesis settings, see \citet{Broekgaarden:2021} for more details.

\begin{table}
\centering
\begin{tabular}{|l|l|l|}
\hline \hline
$\mu$ & Label  & Variation \\ \hline \hline 
A      & fiducial		& --        	     	\\%
\hline
%
B     & $\beta =0.25$  		                 				& fixed mass transfer efficiency of $\beta=0.25$ \\
C       & $\beta =0.5$  	                				& fixed mass transfer efficiency of $\beta=0.5$\\
D        & $\beta =0.75$ 		                 				& fixed mass transfer efficiency of $\beta=0.75$  \\
E       & unstable/no case BB		& case BB mass transfer is always unstable\\
\multirow{2}{*}{F}       & \multirow{2}{*}{ E + K } & case BB mass transfer is always unstable $\&$  \\
 & 	  &   HG donor stars initiating a CE may survive \\
\hline
%
G       &$\alpha=0.1$	                   				&  CE efficiency parameter $\alpha = 0.1$ \\

H       &$\alpha=0.5$	                   				&  CE efficiency parameter $\alpha = 0.5$ \\
I        & $\alpha=2$		                  				& CE efficiency parameter $\alpha = 2$ \\
J       &$\alpha=10$	                   				&  CE efficiency parameter $\alpha = 10$ \\
K      	& optimistic CE	    		& HG donor stars initiating a CE may survive\\
\hline
%
%
L       & rapid SN	                  				& Fryer rapid {SN} remnant mass model \\
M       & $m_{\rm{NS}} = 2\Msun$	                				& maximum NS mass is fixed to $2\Msun$\\
N        & $m_{\rm{NS}} = 3\Msun$		               				& maximum NS mass is fixed to $3\Msun$\\
O       & no PISN		                  				& no  PISN and pulsational-PISN\\
P      	 & $\sigma_{\rm{rms}}^{1D} =100\kms$ 	                				& $\sigma_{\rm{rms}}^{\rm{1D}}=100$\kms  	for core-collapse {SNe}\\
Q        & $\sigma_{\rm{rms}}^{1D} =30\kms$ 	               				& $\sigma_{\rm{rms}}^{\rm{1D}}=30$\kms  	for core-collapse {SNe} \\
R       & $v_{\rm{k,BH}} = 0$	                 				& we assume \acp{BH} receive no natal kick\\
\hline
%
S        &  $f_{\rm{WR}}=0.1$	               				& Wolf-Rayet wind factor $f_{\rm{WR}} = 0.1$ \\
T       &$f_{\rm{WR}}=5$	                 				& Wolf-Rayet wind factor $f_{\rm{WR}} = 5$\\
\hline \hline
\end{tabular}
\caption{
List of the \NmodelsBPS binary population synthesis models studied in this work. 
$\mu$ and `Label' denote the alphabetical letter and abbreviated name used to label each model,  `Variation' denotes what we changed compared to the fiducial settings (model A). 
%
Models B, C, D, E and F vary mass transfer assumptions, models G, H, I, J and K vary common-envelope assumptions, models L, M, N, O, P, Q and R vary supernova assumptions and models S and T vary our Wolf-Rayet wind assumptions. Acronyms used are: common-envelope (CE), Hertzsprung Gap (HG), supernova (SN) neutron star (NS), black hole (BH) and pair-instability SN (PISN). We additionally use the subscripts root-mean-square (rms), one-dimensional (1D) and  Wolf-Rayet (WR). Each model varies one assumption compared to the fiducial model, except for model F where we vary two assumptions (namely those from model E and K; unstable case BB and optimistic CE)}.
\label{tab:variations-BPS}
\end{table}

In models B, C, D, E and F we explore uncertainties in binary mass transfer prescriptions.  Of these, models B, C and D vary the mass transfer efficiency. 
This is defined in \COMPAS by the parameter $\beta$, which determines for a given donated mass rate the fraction that is accreted by the companion star: $\beta \equiv (\diff M_{\rm{acc}} / \diff t) / (- \diff M_{\rm{donor}} / \diff t)$, with $M_{\rm{acc}}$ and $M_{\rm{donor}}$ being the mass of the accretor and donor stars, respectively, and $t$ the time (where $\diff t$ is the simulation time step).  
The excess mass is assumed to leave the binary from the vicinity of the accreting star through `isotropic re-emission'  \citep[e.g.,][]{1975MmSAI..46..217M,1991PhR...203....1B,1997A&A...327..620S} and the angular momentum loss is calculated accordingly (cf., Equations~32 and 33 in \citealt{2008ApJS..174..223B}).  
Our fiducial model (A) assumes the accretion rate  is limited by  the star's thermal  timescale: $\diff  M_{\rm{acc}} / \diff t \leq 10 M_{\rm{acc}} / \tau_{\rm{KH}}$, where $\tau_{\rm{KH}}$ is the Kelvin-Helmholtz (thermal) timescale\footnote{Given in Equation 61 of \citet{2002MNRAS.329..897H}, where we use a pre-factor of 30 Myr from Equation 2 in \citet{KalogeraWebbink:1996}.}, and the factor of 10 is added to take into account the expansion of the accretor due to mass transfer (cf., \citealt{1972AcA....22...73P}, \citealt{2002MNRAS.329..897H} and \citealt{2015ApJ...805...20S}).  
Models B, C and D assume a different accretion rate limit for stars by fixing $\beta$ to $0.25$, $0.5$, and $0.75$, respectively. 
All of our models assume an Eddington-limited accretion rate for compact objects. In our fiducial model we assume that case BB mass transfer, which is mass transfer from a stripped post-helium-burning star onto the accretor \citep{1981A&A....96..142D}, is always stable. 
Model E explores a variation where case BB (and case BC, but these are more rare) mass transfer is assumed to always be unstable, leading to a \ac{CE} phase.  This assumption causes case BB mass transferring systems to merge as stars in model E as our fiducial model assumes the now unstable \ac{CE} phase initiated by a (helium) Hertzsprung Gap star is unsuccessful and leads to a merger (as described further in the `pessimistic \ac{CE} assumption' in the next paragraph).  
We therefore also add model F, which explores the effect of assuming unstable case BB mass transfer, but allowing Hertzsprung Gap donor stars to survive a \ac{CE} phase (the `optimistic \ac{CE} assumption').

In models G, H, I, J and K we explore the effect of changing the \ac{CE} prescription assumptions, which in \COMPAS are parameterized using the `$\alpha$--$\lambda$' formalism from \citet{1984ApJ...277..355W} and \citet{1990ApJ...358..189D}. Our fiducial model assumes a \ac{CE} efficiency of $\alpha_{\rm{CE}}=1$ and uses for $\lambda$ the ``Nanjing lambda'' prescription   \citep[cf.,][]{2012ApJ...759...52D}, which is based on models from \citet{2010ApJ...716..114X,2010ApJ...722.1985X}. In models G, H, I  and J we change the \ac{CE} efficiency to fixed values of $0.1$,  $0.5$, $2$ and $10$, respectively.  Compared to the fiducial model, lower and higher values of $\alpha_{\rm{CE}}$ reduce and increase the efficiency with which the \ac{CE} is ejected, respectively.  In model K (and model F) we allow Hertzsprung gap stars that initiate a \ac{CE} to possibly survive the \ac{CE} (also known as the `optimistic' \ac{CE} assumption), whereas in our fiducial model these are assumed to always lead to an unsuccessful \ac{CE} ejection (and merger), the `pessimistic' \ac{CE} scenario \citep[cf.,][]{2012ApJ...759...52D}. 

In models L, M, N, O, P, Q and R we vary the \ac{SN} prescription assumptions. In model L we use the `rapid' \ac{SN} remnant mass model of \citet{2012ApJ...749...91F} instead of their `delayed' model, which is implemented in our fiducial model. The rapid model creates a mass gap between $\approx2-6$\Msun, where no \acp{BH} form, whereas in the delayed model such \acp{BH} can form. Models M and N change our assumption for the maximum \ac{NS} mass, by default 2.5 \Msun, to $2\Msun$ and $3\Msun$, respectively and we adapt the remnant mass prescription from \citet{2012ApJ...749...91F} accordingly. In model O we do not use the prescription for pair-instability \acp{SN} and pulsational pair-instability \acp{SN}, therefore allowing the formation of \acp{BH} in the mass range of  $\approx 40{-}100$\Msun. In models P and Q we change the root mean square velocity dispersion (\sigmacc) for the Maxwellian \ac{SN} natal kick distribution for both \acp{BH} and \acp{NS}, to 100\kms and 30\kms, respectively (the fiducial model uses  $\sigmacc=265\kms$). For all our models we assume that a fraction of the ejected material ($f_{\rm{fb}}$) falls back onto the compact object and we adjust the remnant mass and re-scale the \ac{SN} kick magnitude accordingly (cf., \citealt{2012ApJ...749...91F}). For ultra-stripped \acp{SN} and electron-capture \acp{SN} we always draw the \ac{SN} kick using a one-dimensional root-mean-square velocity dispersion of  $\sigma_{\rm{rms}}^{1D} = 30$\kms  following \citet{2002ApJ...571L..37P} and \citet{2004ApJ...612.1044P}. In model R we assume instead that only all \acp{BH} receive zero natal kicks, $\vk=0$\kms. 

Finally, models S and T explore the assumptions for the mass loss rate in Wolf-Rayet winds. Our Wolf-Rayet wind prescription follows \citet{Belczynski:2010}, which is based on \citet{1998A&A...335.1003H} and  \citet{2005A&A...442..587V}, by parameterizing the wind strength with a multiplicative parameter $f_{\rm{WR}}$ (cf., \citealt{2018MNRAS.477.4685B}). By default we use $f_{\rm{WR}} = 1$ and in models S and T we vary this to $0.1$ and $5$, respectively, which largely spans the possible range for Wolf-Rayet winds inferred from observations (e.g., \citealt{Vink:2017,2019A&A...625A..57H,2019A&A...627A.151S,Sander:2020}). 

\subsection{Metallicity-dependent star formation rate density models}
\label{sec:method-metallicity-specific-star-formation-rate-density-prescription}

The time between formation and merger of a \ac{DCO} can range up to many Gyr \citep[e.g.,][]{Neijssel:2019}. 
As a result, the \acp{DCO} that are detected by current ground-based \ac{GW} observatories can originate from stars that formed throughout a wide range of redshifts with a large variety of birth metallicities \citep[e.g.,][]{Chruslinska:2019obsSFRD}. Similar to  \citetalias{Broekgaarden:2021}, we follow the method of \citet{Neijssel:2019} to create a metallicity-dependent star formation rate density \SFRD, which describes the star formation history as a function of birth (initial) metallicity\footnote{Where Z is the fractional metallicity, such that $\rm{X}+\rm{Y}+\rm{Z}=1$ with X and Y the mass fractions of hydrogen and helium, respectively.} \Zi and redshift $z$. To explore uncertainties in \SFRD, we use a total of \NmodelsMSSFR \SFRD models. Each model convolves a \ac{SFRD}\footnote{We use \SFRD for the metallicity-dependent star formation rate density as a function of metallicity and redshift and \ac{SFRD} for the total star formation rate density across all metallicities, which we model as only a function of redshift.} with a metallicity probability distribution function, $\diff P / \diff \Zi $.  Our first model $\rm{xyz}=000$ (see Table~\ref{tab:MSSFR-variations-labels} for the labeling) is the so-called  `preferred' phenomenological model from \citet{Neijssel:2019}, which they fit to match the \ac{BHBH} observations from the first two observing runs by LIGO and Virgo.  
The other 27 ($3\times 3\times 3$) models are constructed by combining a \ac{SFRD} with a \ac{GSMF}, and a \ac{MZR}. The three variations for each \ac{SFRD}, \ac{GSMF} and \ac{MZR} that we consider are given in Table~\ref{tab:MSSFR-variations-labels}. Although many of the prescriptions in these models oversimplify the complex \SFRD that is evident from observations, these prescriptions provide a convenient parametrization for binary population synthesis studies. For more details about caveats and uncertainties in the modelling of \SFRD within binary population synthesis we refer the reader to \citet{Chruslinska:2019obsSFRD}, \citet{Neijssel:2019}, \citet{Boco:2021} and references therein.


\section{Results}
\label{section:results}

\subsection{Metallicity specific merger yields}
\label{section:Formation-rates-per-metallicity}
\begin{figure*}
    \centering
\includegraphics[width=1.0\textwidth]{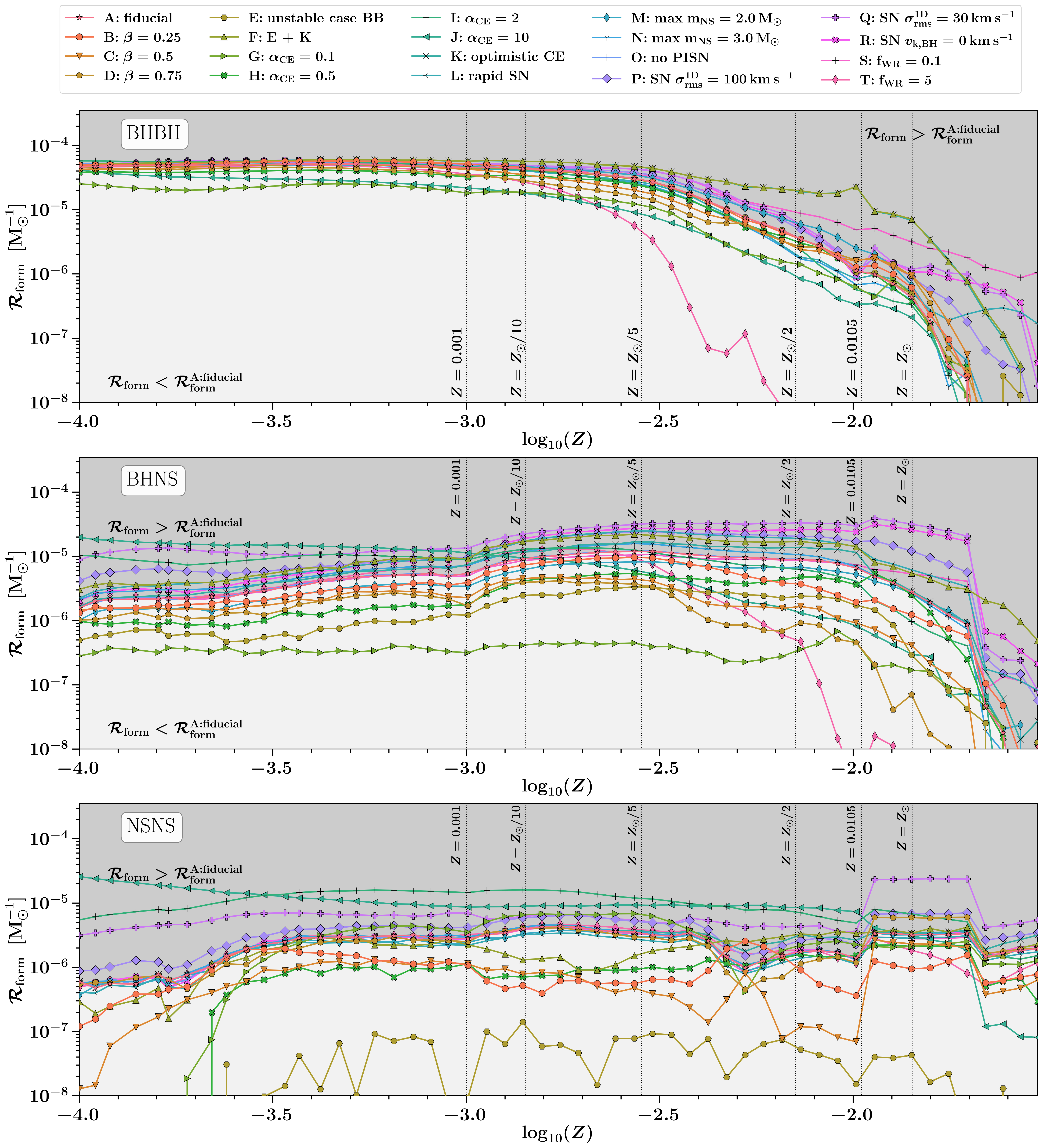}
\caption{Formation yield of merging double compact objects  per solar mass of star formation, $\mathcal{R}_{\rm{form}} = \diff N_{\rm{form}} / \diff \MSFR$ (Equation~\ref{eq:formation-rate-COMPAS}), as a function of birth metallicity \Zi. The yield only includes binaries that have \ac{GW}-driven inspiral times  $\lesssim 14\Gyr$. 
From top to bottom the panels show the formation yields for merging \ac{BHBH}, \ac{BHNS} and \ac{NSNS}, respectively. Each color and marker type correspond to one of the \NmodelsBPS binary population synthesis models explored in this study (Table~\ref{tab:variations-BPS}). The dark and light gray areas in the background mark where the formation yield is larger and smaller compared to our fiducial model (A) yield. Vertical dotted lines show fixed \Zi values to guide the reader. The marker points show the \Zi grid values that we simulated with {\COMPAS} in each simulation. The sharp increase in yield around $\Zi \approx 0.0105$, particularly visible in the \ac{NSNS} panel, is caused by an artificial bifurcation in our $\lambda$ values for the common-envelope treatment in \COMPAS at this \Zi. The scatter in the unstable case BB mass transfer model (E) in the \ac{NSNS} panel is caused by sampling noise. \href{https://github.com/FloorBroekgaarden/Double-Compact-Object-Mergers/blob/main/plottingCode/Fig_1_and_Fig_A1/FormationRateAllModels3panels_vertical.pdf}{\faFileImage} \href{https://github.com/FloorBroekgaarden/Double-Compact-Object-Mergers/blob/main/plottingCode/Fig_1_and_Fig_A1/make_Fig_1_and_Fig_A1.ipynb}{\faBook}}
    \label{fig:FormationRateDCO-per-metallicity}
\end{figure*}

In Figure~\ref{fig:FormationRateDCO-per-metallicity} we show the calculated formation yields ($\rate_{\rm{form}}$) for \ac{BHBH}, \ac{BHNS} and \ac{NSNS} mergers per unit solar mass of stars formed ($\diff \MSFR$) as a function of \Zi for our \NmodelsBPS binary population synthesis models. This formation yield only includes \ac{DCO} systems that have a \ac{GW} driven inspiral time (\tinspiral) that is smaller than the Hubble time ($\approx 14\Gyr$) and is given by
\begin{align}
\rate_{\rm{form}}(\Zi) = 
 \frac{\diff \Nform}{\diff \MSFR } (\Zi), 
\label{eq:formation-rate-COMPAS}
\end{align}
where we convert the number of mergers in our COMPAS realizations to formation yields by integrating our simulation range over the full initial parameter space of binary systems (e.g., the full initial mass range of stars) and assuming a corrected\footnote{Corrected to take into account the differences in parameter range between population synthesis and observations, see \citet{deMinkBelczynski:2015}.} binary fraction of $f_{\rm{bin}}=1$ \citep[consistent with e.g.,][]{2017IAUS..329..110S}.  Details are given in  \citetalias{Broekgaarden:2021}. 

\subsubsection{Impact from the binary massive star evolution assumptions and metallicity}

Figure~\ref{fig:FormationRateDCO-per-metallicity} demonstrates that the variation in the \ac{BHBH} merger formation yield is typically dominated by \Zi, causing variations on the order of $\mathcal{O}(10^2)$ between $\Zi\approx 0.001$ and $\Zi \approx \Zsun$. Variations in our massive binary population synthesis assumptions impact the \ac{BHBH} merger yield only with factors of up to $\approx 3$ at $\Zi \lesssim 0.1\Zsun$ and on the order of $\mathcal{O}(10)$ for most stellar evolution models at higher \Zi. The exceptions to this include model T ($f_{\rm{WR}}=5$), which has a much lower BHBH yield (by $\mathcal{O}(10^2)$) compared to other stellar evolution models for $\Zi \gtrsim 0.2\Zsun$ where BH formation is drastically reduced due to the strong stellar wind mass loss. Other examples include the optimistic \ac{CE} models (F and K) that have higher  \ac{BHBH} yields compared to other stellar evolution models for $\Zi \gtrsim 0.2\Zsun$. 

The scatter in the \ac{BHNS} and \ac{NSNS} formation yields, on the other hand, are typically dominated by variations in the stellar and binary evolution models over most of the \Zi range, leading to variations of $\gtrsim \mathcal{O}(10^2)$ in the rates, compared to typical ranges of $\lesssim \mathcal{O}(10)$ from variations in \Zi for most models.

\subsubsection{Trends in the metallicity-dependent formation yields}
As can be seen in Figure~\ref{fig:FormationRateDCO-per-metallicity}, the formation yield of \ac{BHBH} mergers is constant at low metallicities and then steeply declines at $\Zi \gtrsim 0.1 \Zsun$ for all model variations.  This steep decline of merging BHBHs at higher metallicities has been pointed out previously \citep[e.g.,][]{Belczynski:2010,2012ApJ...759...52D}, and is consistent with various recent works including  \citet{2018MNRAS.474.2959G}, \citet{GiacobboMapelli:2018}, \citet{2018A&A...619A..77K}, \citet{Chruslinska:2019obsSFRD}, \citet{Neijssel:2019} and \citet{Santoliquido:2021}, demonstrating that this trend is present among many different population synthesis studies and simulation settings. This decline is due to several metallicity dependent effects. First, stronger stellar winds at higher metallicities lead to increased mass loss and hence wider binaries with less massive \acp{DCO}, reducing the fraction of \acp{BHBH} that will merge in a Hubble time. In addition, the less massive \acp{BH} are assumed to have smaller amounts of fallback and hence receive larger natal kicks in our simulations, increasing the fraction that is disrupted during the \ac{SN} \citep[cf.,][]{2012ApJ...749...91F,2018A&A...619A..77K,2019A&A...624A..66R}.  Indeed, in the model variations where \acp{BH} receive lower or no \ac{SN} kick (Q and R) or the Wolf-Rayet winds are weaker (model S) the decline in \ac{BHBH} yield is less severe in Figure~\ref{fig:FormationRateDCO-per-metallicity}.  Second, there is a metallicity dependence of the radius expansion of stars in our single star evolution tracks \citep{2000MNRAS.315..543H}, leading more stars to enter the \ac{CE} phase during the Hertzsprung Gap instead of during the giant phase at higher ($\sim{\rm Solar}$) metallicities  (cf., Figure~2 in \citealt{Belczynski:2010}, Figure~7 in \citealt{Linden:2010} and Section 3.1.1 in \citealt{Bavera:2021}). In our fiducial simulation such Hertzsprung Gap donor stars do not survive the \ac{CE} phase. In models F (E+K) and K (optimistic CE) we instead assume these systems do survive, leading to a visible boost in BBH yields at higher $\Zi$, as seen in Figure~\ref{fig:FormationRateDCO-per-metallicity}. 

The behavior of the \ac{BHNS} yield in most models, on the other hand, first modestly increases as a function of \Zi and then declines around $\Zsun$ (Figure~\ref{fig:FormationRateDCO-per-metallicity}; cf., \citealt[][]{2018A&A...619A..77K,Chruslinska:2019obsSFRD,RomanGarza:2020}). This causes the yield to broadly peak in the range $0.2\Zsun\lesssim \Zi \lesssim \Zsun$, where the location of the peak varies between models as shown in Figure~\ref{fig:FormationRateDCO-per-metallicity}. The same metallicity dependent effects as described for \ac{BHBH} mergers above are at play for \ac{BHNS} mergers. However, since \ac{BHNS} systems typically already form \acp{DCO} with lower mass \acp{BH} and form from lower mass stars compared to \ac{BHBH} mergers, these \ac{BHNS} systems do not experience the same boost in formation yields at low metallicities compared to the \ac{BHBH} mergers, leading to lower yields at lower \Zi. 
In addition, the peak in \ac{BHNS} yield is caused by the maximum radial extent of stars in the Hertzsprung Gap, which generally increases  with \Zi but decreases  between $ 10^{-3} \lesssim \Zi \lesssim 10^{-2}$ in our simulations for the zero-age main sequence mass range of \ac{BHNS} progenitors (initial primary masses of $15 \lesssim \monei / \Msun \lesssim 50 $). This is based on the single star evolution prescriptions from \citet{2000MNRAS.315..543H} implemented in {\sc{COMPAS}}. As a result of the smaller radii there are typically fewer stellar mergers in our simulations as fewer stars completely engulf their companion during mass transfer \citep[as discussed  by][]{GiacobboMapelli:2018} and the average initial separations of binaries that form merging \acp{BHNS} shifts to smaller values, which are more commonly formed. See also the discussion in \S4.1.2 in \citet{2018A&A...619A..77K} for additional details.

We find that the \ac{NSNS} merger yield remains roughly constant as a function of \Zi \citep[cf.,][]{Chruslinska:2018,2018A&A...619A..77K, GiacobboMapelli:2018,Neijssel:2019} as the metallicity-dependent effects described above are less significant for lower-mass stars that form \ac{NSNS} binaries. The strong increase in formation yield around $\Zi \approx 0.0105$ in Figure~\ref{fig:FormationRateDCO-per-metallicity} that is particularly noticeable for \ac{NSNS} binaries is due to a bifurcation in our $\lambda$ prescription for the envelope binding energy (relevant for the \ac{CE} phase) and is an artifact in our modelling.  More orbital energy is needed to successfully unbind the envelope for $\Zi >0.0105$ in our models. This means that the binary engulfed in the \ac{CE} will need to move to tighter orbits in order to successfully eject the envelope. This results in tighter post-\ac{CE} binaries that survive the second \ac{SN} and can merge in a Hubble time, thereby increasing the rate. 
Finally, we note that the scatter in model E (unstable case BB) is caused by sampling noise as a consequence of the very small number of \ac{NSNS} mergers (only 371) in this simulation\footnote{See our online table \href{https://raw.githubusercontent.com/FloorBroekgaarden/Double-Compact-Object-Mergers/main/otherFiles/DCO_table_detailed.png}{\faFileImage}.}. In the next Section we discuss this model and  show that it under-predicts the inferred NSNS rate from \ac{GW} observations by two orders of magnitude. 

Comparing the panels in Figure~\ref{fig:FormationRateDCO-per-metallicity} we find that at low \Zi the formation yield of merging \acp{BHBH} in a Hubble time exceeds that of merging \acp{BHNS} and \acp{NSNS}. The merging \ac{BHNS} formation yield starts dominating over the \ac{BHBH} yield at $\Zi \gtrsim 0.3\Zsun$, whereas for merging \ac{NSNS} this occurs at $\Zi \gtrsim 0.5\Zsun$ for most models.   We show these merging \ac{DCO} ratios in  Figure~\ref{fig:FormationRateDCO-Ratios-per-metallicity}. Using our \SFRD models we can translate these transition metallicities into approximate typical redshifts. For example, Figure~\ref{fig:2panels_examples_SFRD-Z} shows that the average metallicity of star formation is  $\langle \Zi \rangle \approx 0.5\Zsun$ at $z\approx 3$ and $z\approx 0.4$ for the $\rm{xyz}=312$ and $\rm{xyz}=231$ \SFRD models, respectively. These two \SFRD models correspond to the highest and lowest average metallicity of star formation within our simulated \SFRD models and, as we will show in the next section, to one of the lowest and highest \ac{DCO} merger rate density predictions, respectively.

\subsection{Intrinsic merger rates}
\label{sec:results-comparing-BBH-BNS-intrinsic-rates}

\begin{figure*}
    \centering
\includegraphics[width=1.0\textwidth]{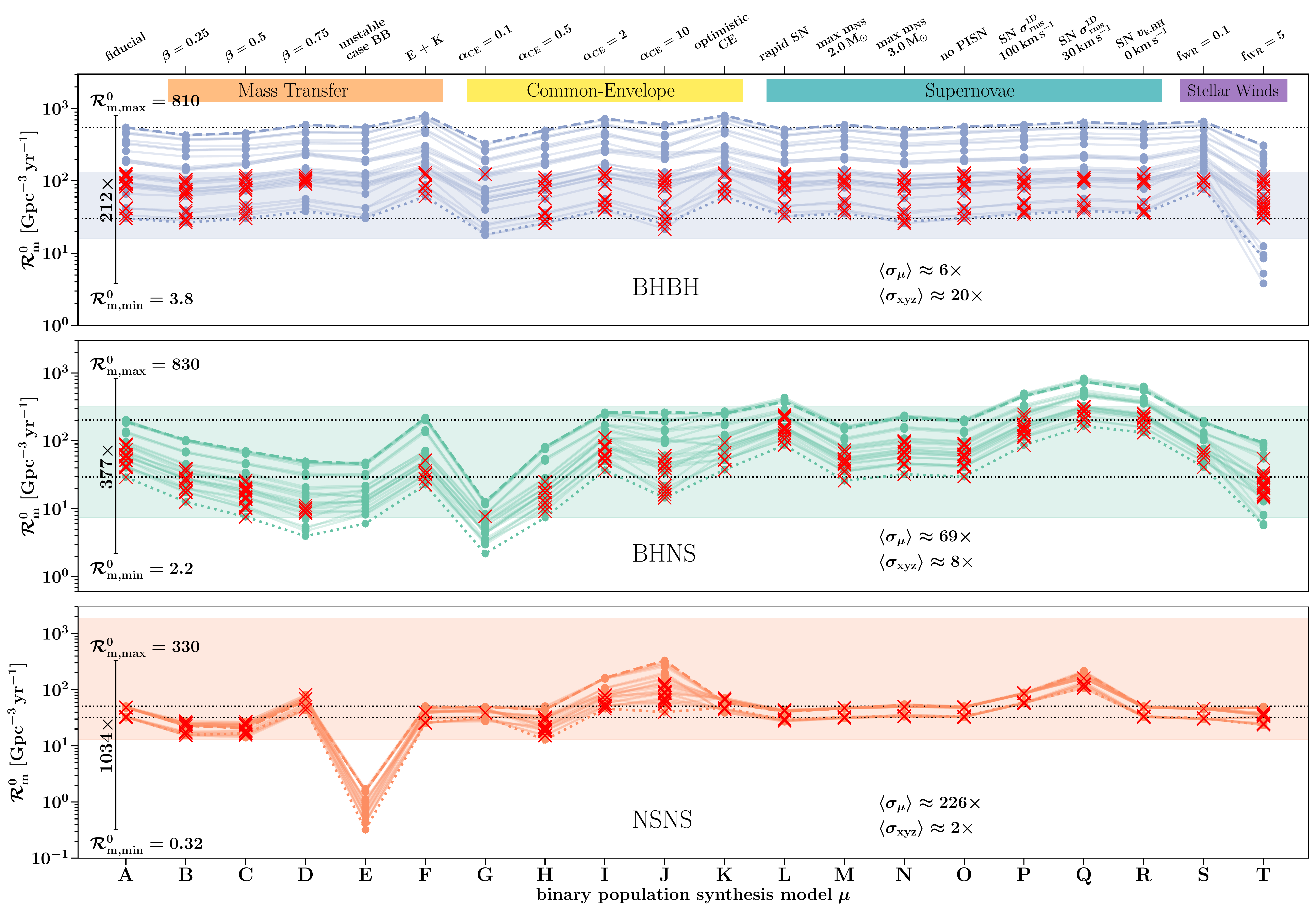} 
    \caption{Intrinsic (local) merger rate densities for merging \acp{BHBH} (top panel), \acp{BHNS} (middle panel) and \acp{NSNS} (bottom panel) systems for our \Nmodels model variations. The rates are for merging \acp{DCO} at $z\approx0$ calculated using Equation~\ref{eq:MSSFR-merger-rate} (i.e., without applying \ac{GW} detector selection effects).    We show for each of the \NmodelsBPS binary population synthesis models (Table~\ref{tab:variations-BPS}) the merger rates for the \NmodelsMSSFR variations in \SFRD (Table~\ref{tab:MSSFR-variations-labels}). We connect predictions that use the same \SFRD model with a line for visual reasons only. Two \SFRD variations are highlighted  corresponding to \SFRD models resulting in one of the highest  (\rm{xyz}$=$231, dashed) and lowest  (\rm{xyz}$=$312, dotted) merger rate predictions. The values $\langle \sigma_{\mu}\rangle$ and $\langle \sigma_{\rm{xyz}}\rangle$ quote a proxy for the mean scatter (Equations~\ref{eq:sigma-mu} and~\ref{eq:sigma-xyz}) in the predicted rates due to variations in binary population synthesis and \SFRD, respectively.      The minimum and maximum rates and their ratio  are quoted with a black error bar on the left.  We use the short-hand notation $\rate_{\rm{m}}^0 \equiv (\diff \Ndet^2 / \diff \ts \diff \Vc)(\tmerger(z\approx0))$.      Shaded horizontal bars indicate the ranges spanned by the  $90\%$ credible intervals for the intrinsic merger rates inferred from GW observations  \citet{Abbott:2021GWTC3pop}:   $\rate_{\rm{m}}^{0,\rm{BHBH}}= \RateGWTCminBHBH{-}\RateGWTCmaxBHBH$\GpcminThree\yearmin  (\ac{BHBH}),  $\rate_{\rm{m}}^{0,\rm{BHNS}} = \RateGWTCminBHNS{-}\RateGWTCmaxBHNS$\GpcminThree\yearmin (\ac{BHNS}) and  $\rate_{\rm{m}}^{0,\rm{NSNS}}  =  \RateGWTCminNSNS{-}\RateGWTCmaxNSNS$\GpcminThree\yearmin  (\ac{NSNS}).    Model realizations where all three of the \ac{BHBH}, \ac{BHNS} and \ac{NSNS} simulated merger rate densities overlap with the inferred ranges from  observations are marked with red crosses. 
    A video showing the rates for each individual \SFRD model is provided at \url{https://youtu.be/LkT7FD1xC2g}. 
    Dashed horizontal lines indicate the range for the fiducial model (A). At the top we added colored labels to indicate what physics assumptions are varied compared to our fiducial assumptions in the models. \href{https://github.com/FloorBroekgaarden/Double-Compact-Object-Mergers/blob/main/plottingCode/Fig_2/Rates_intrinsic_with_GWTC2-2.pdf}{\faFileImage} \href{https://github.com/FloorBroekgaarden/Double-Compact-Object-Mergers/blob/main/plottingCode/Fig_2/make_Figure_2.ipynb}{\faBook}
    }%
    \label{fig:IntrinsicRatesBBHBNSBHNS}
\end{figure*}

In Figure~\ref{fig:IntrinsicRatesBBHBNSBHNS} we show the population synthesis calculated local (intrinsic) merger rate densities for \ac{BHBH}, \ac{BHNS} and \ac{NSNS} binaries. There are three main findings that we describe in more detail below. First, we find that the combined variations in our massive binary star and \SFRD assumptions impact the intrinsic \ac{DCO} rates with resulting ranges of $\approx 210{-}1030$. Second, only a subset of the \Nmodels models matches the ranges for the inferred \ac{BHBH}, \ac{BHNS} and \ac{NSNS} merger rate density from \ac{GW} observations \citep{Abbott:2021GWTC3pop}. Third, we find that the calculated merger rates for different types of \ac{DCO} are sensitive to different uncertainties in the modelling.  Most strikingly, we find that the \ac{BHBH} merger rates are primarily sensitive to variations in \SFRD models, whereas the \ac{NSNS} merger rates are predominantly impacted by binary stellar evolution model variations. This means that the observed \ac{BHBH} and \ac{NSNS} rates can provide a test-bed for \SFRD and stellar evolution uncertainties, respectively. 

The intrinsic rates are calculated using the formation yields from Equation~\ref{eq:formation-rate-COMPAS} and by taking into account the \SFRD weighting discussed in \S\ref{sec:method-metallicity-specific-star-formation-rate-density-prescription}. The local merger rate is then obtained using $\tmerger(z\approx0)$ in the equation (i.e. evaluating the equation at the center of our lowest redshift bin for $z$)
\begin{align}
&\rate_{\rm{m}} (\tmerger) = \frac{\diff^2 \Nmerger }{\diff \tmerger \diff \Vc } (\tmerger) = \notag \\
		&\int \diff \Zi  \int_0^{\tmerger} \diff \tdelay \, {\mathcal{S}}(\Zi, \tform) \,  \frac{\diff^2 \Nform}{\diff \MSFR  \diff \tdelay} (\Zi, \tdelay),  
\label{eq:MSSFR-merger-rate}
\end{align}
where \tmerger again is time in the source frame measured from the Big Bang\footnote{Where we use the WMAP9--cosmology from Astropy \citep{2013ApJS..208...19H}. This assumption does not drastically impact our results.}, \tmerger is the time of the \ac{DCO} merger, \tdelay is the delay time between the formation of the \ac{DCO} and its merger, and \Vc is the co-moving volume. The delay time is $\tdelay = \tevolve+ \tinspiral$, the total time from the onset of hydrogen burning at ZAMS to forming a \ac{DCO}, i.e., until the second SN (\tevolve), and the time it takes the \ac{DCO} to coalesce from the moment of the second \ac{SN} (\tinspiral); see Figure~1 in  \citetalias{Broekgaarden:2021} for more details. The binary initially forms at $\tform$, which we set as $\tform = \tmerger - \tdelay$ in this equation.  For the \SFRD we use the \NmodelsMSSFR models from Table~\ref{tab:MSSFR-variations-labels}. In practice we estimate Equation~\ref{eq:MSSFR-merger-rate} with a Riemann sum where we sum over our metallicity grid and redshift (time) bins.  

To quantify the scatter in the predicted intrinsic rates we calculate the mean of the ratios between the maximum and minimum predicted rates given by
\begin{equation}
\label{eq:sigma-mu}
\langle \sigma_{\rm{\mu}} \rangle = \frac{1}{28} \sum_{\rm{xyz}=000}^{\rm{xyz}=333}
 \frac{\rm{max}(\rate_{\rm{m, Axyz}}^0,... ,\rate_{\rm{m, Txyz}}^0 )}{\rm{min}(\rate_{\rm{m, Axyz}}^0,... ,\rate_{\rm{m, Txyz}}^0)}, 
\end{equation} 
and 
\begin{equation}
\label{eq:sigma-xyz}
\langle \sigma_{\rm{xyz}}  \rangle = \frac{1}{20} \sum_{\mu=A}^{\mu=T}
 \frac{\rm{max}(\rate_{\rm{m, \mu000}}^0,... ,\rate_{\rm{m, \mu333}}^0 )}{\rm{min}(\rate_{\rm{m, \mu000}}^0,... ,\rate_{\rm{m, \mu333}}^0 ) }, 
\end{equation} 
where we used the short-hand notation $\rate_{\rm{m}}^0  = \rate_{\rm{m}}(\tmerger(z\approx0))$ and $\mu$ and $\rm{xyz}$ are the stellar evolution and \SFRD labels, respectively.  Intuitively,  Equation~\ref{eq:sigma-mu} represents the uncertainty range from stellar evolution as averaged over the \SFRD models, whereas Equation~\ref{eq:sigma-xyz} represents the uncertainty range from \SFRD as averaged over the stellar evolution models. Large (small) values for $\sigma_{\mu}$ and $\sigma_{\rm{xyz}}$ correspond to large (small) impacts by binary star evolution and \SFRD variations, respectively.

\subsubsection{BHBH merger rates}

We find that the intrinsic merging \ac{BHBH} rates are predicted to lie in the range  $\rate_{\rm{m}}^{0,\rm{BHBH}}\approx \RateIntrinsicAzeroBHBHmin{-} \RateIntrinsicAzeroBHBHmax$\GpcminThree \yearmin as shown in the top panel in Figure~\ref{fig:IntrinsicRatesBBHBNSBHNS}.  The combined uncertainty thus impacts the predicted rates with a factor of up to\footnote{This is the ratio between the minimum and maximum predicted rates quoted with the error bar in the left of each panel in Figure~\ref{fig:IntrinsicRatesBBHBNSBHNS}. This is not the same as multiplying $\langle \sigma_{\rm{xyz}} \rangle$ and $\langle \sigma_{\mu} \rangle$, as the latter are averages calculated using Equations~\ref{eq:sigma-mu} and \ref{eq:sigma-xyz}, and because the impacts from stellar evolution and \SFRD variations are not (fully) independent.} $\approx 210$. We find that the variations in stellar and binary evolution impact the \ac{BHBH} rate on average with $\langle\sigma_{\mu} \rangle\approx  6$, while variations in our \SFRD models impact the predicted \ac{BHBH} rate with uncertainties of $\langle \sigma_{\rm{xyz}}\rangle \approx 20$. 

We find that the majority of our stellar and binary evolution model variations do not impact the \ac{BHBH} rate with more than a factor $\approx 2$, as can be seen in the top panel of Figure~\ref{fig:IntrinsicRatesBBHBNSBHNS} when comparing the predicted rates with our fiducial model. Model T ($f_{\rm{WR}}=5$) has the highest impact on the rates. The other largest \ac{BHBH} rate changes are from stellar evolution models that change the \ac{CE} assumptions (e.g., models F, G and K). 

The $90\%$ credible intervals from the \ac{GW} inferred rate of merging \acp{BHBH} spans $\rate_{\rm{m}}^{\rm{BHBH}} =  \RateGWTCminBHBH{-}\RateGWTCmaxBHBH$\GpcminThree\yearmin based on the {GWTC-3} catalog from \citet{Abbott:2021GWTC3pop}\footnote{From hereon we quote the rate range from the `merged' row in Table II in \citet{Abbott:2021GWTC3pop} that represents the union of 90$\%$ credible intervals from their PDB, MS and BGP models.}.  
Figure~\ref{fig:IntrinsicRatesBBHBNSBHNS} shows that only a subset  of our \SFRD models (lower lines in the top panel) predict merging \ac{BHBH}  rates that are consistent with this range. 
The \SFRD realizations that typically over-predict the observed \ac{BHBH} rate for all stellar evolution models are the ones with an implementation of the \citet{2006ApJ...638L..63L} \ac{MZR} $(\rm{z}=1)$. This \ac{MZR} model corresponds to the lowest average \Zi  among our \ac{MZR} variations (right panel, Figure~\ref{fig:2panels_examples_SFRD-Z}), which significantly increases the \ac{BHBH} yield. The over-prediction is even more significant if other formation channels than the isolated binary evolution channel further contribute to the observed \ac{BHBH} rate \citep[e.g.,][]{Abbott:2021GWTC2pop,Zevin:2021-branching-ratios}. 

\subsubsection{BHNS merger rates}
\label{subsec:intrinsic-merger-rates-BHNS}

For the merging \ac{BHNS} systems we find rates in the range $\rate_{\rm{m}}^{0,\rm{BHNS}} \approx  \RateIntrinsicAzeroBHNSmin{-}\RateIntrinsicAzeroBHNSmax$\GpcminThree \yearmin as shown in the middle panel of Figure~\ref{fig:IntrinsicRatesBBHBNSBHNS}.   Our combined model variations impact the predicted \ac{BHNS} rates with an uncertainty factor of up to $\approx 377$. We find that variations in the rate are typically $\langle \sigma_{\mu} \rangle \approx 69$ for binary evolution variations, while the impact from \SFRD variations typically leads to a range of $\langle \sigma_{xyz}\rangle\approx 8$. 

Almost all predicted \ac{BHNS} rates are consistent with the range $\rate_{\rm{m}}^{0,\rm{BHNS}} =  \RateGWTCminBHNS{-}\RateGWTCmaxBHNS$\GpcminThree\yearmin spanned by the $90\%$ credible intervals from observations  \citep{Abbott:2021GWTC3pop}. The exceptions are a few of the \SFRD models in combination with models D, E and G that under-predict the \ac{BHNS} rate and a few of the \SFRD models with $z=1$ in combination with stellar evolution model P, Q and R (lower \ac{SN} kicks),  which slightly over-predict the observed upper limit.  The latter models have reduced \acp{SN} kicks, which increases the fraction of systems that  stay bound during the \ac{SN} compared to our fiducial model (see  \citetalias{Broekgaarden:2021} for a more detailed discussion). Future improved constraints might enable ruling out models.

\subsubsection{NSNS merger rates}

The bottom panel in Figure~\ref{fig:IntrinsicRatesBBHBNSBHNS} shows our predicted merging \ac{NSNS} rates.  We find values in the range $\rate_{\rm{m}}^{0,\rm{NSNS}}\approx  \RateIntrinsicAzeroNSNSmin{-}\RateIntrinsicAzeroNSNSmax$\GpcminThree \yearmin. Combined, our \Nmodels model realizations impact the estimated \ac{NSNS} rates up to a factor of $\approx 1030$. The uncertainty from stellar and binary evolution dominates the scatter in the calculated \ac{NSNS} rates (cf., \citealt{Santoliquido:2021}), impacting the rate by $\langle\sigma_{\rm{\mu}}\rangle \approx 226$ compared to $\langle \sigma_{\rm{xyz}}\rangle \approx 2$ when varying the \SFRD.  

All of our calculated \ac{NSNS} merger rates except those involving model E match the observed merger rate density range of $\rate_{\rm{m}}^{0,\rm{NSNS}} =\RateGWTCminNSNS{-} \RateGWTCmaxNSNS$\GpcminThree\yearmin \citep{Abbott:2021GWTC3pop}, although the majority of rates fall in the lower end of the observed \ac{NSNS} merger rate range.   
Models I, J, P and Q have the highest predicted \ac{NSNS} merger rate densities as the \ac{NSNS} rate is boosted in these models. For example, in models I and J fewer binaries merge during the \ac{CE} phase compared to the fiducial model. In models P and Q the smaller value for the root-mean-square velocity dispersion reduces the kick velocities that decreases the number of binaries that disrupts during the \ac{SN}. Only model E is in particular an outlier in the predicted \acp{NSNS} merger rate. In this channel the formation yield of merging \acp{NSNS} is extremely low (Figure~\ref{fig:FormationRateDCO-per-metallicity}). This is because the majority of systems leading to merging \acp{NSNS} experience a case BB mass transfer phase that ultra-strips the NS progenitor \citep{Dewi:2003,2017ApJ...846..170T}. In our model variation E we assume this phase of mass transfer to be unstable, leading to a stellar merger in combination with our pessimistic \ac{CE} assumption. Indeed, in model F where we have the same settings as in model E but add the `optimistic' CE assumptions, we obtain \ac{NSNS} rates consistent with our fiducial model. If we exclude model E we find instead  $\langle \sigma_{\rm{\mu}}\rangle \approx 10 $ (while  $\langle \sigma_{\rm{xyz}}\rangle$ remains unchanged).

\subsubsection{Trends with stellar evolution variations}
There are a few trends visible in the rates in Figure~\ref{fig:IntrinsicRatesBBHBNSBHNS}. First, the predicted \ac{BHNS} rate declines with increasing mass transfer efficiency, $\beta$ (models B, C and D). The increase in $\beta$ leads the secondary to accrete more mass during the first stable mass transfer phase, resulting in more massive secondaries and typically larger separations because the mass transfer is more conservative.  
The larger masses and/or larger separations at higher $\beta$ values results in  more binaries eventually disrupting during a \ac{SN}, merging during the \ac{CE} phase, and/or forming a \ac{BHBH} binary instead of a BHNS (cf., \citealt{Kruckow:2018}). In addition, this  also impacts \tinspiral in a complex way due to the interplay of the typically larger separation after the first stable mass transfer phase, as well as more orbital shrinking during the \ac{CE} phase because the star has to eject a more massive envelope.    

The \ac{BHBH} and \ac{NSNS} rates instead increase with increasing values for $\beta$.  This different behavior comes from a non-trivial combination of ways in which the formation pathways of binaries leading to merging \acp{BHNS} are different from merging \acp{BHBH} and \acp{NSNS} and how changing $\beta$ impacts this.  An example is that the impact from $\beta$ on the evolution of the binary is connected with the mass ratios of the binary (e.g. the mass ratio impacts the size of the Roche lobes, whether the  mass transfer is stable or unstable and whether the mass transfer causes the binary orbit to widen or shrink). The merging \ac{BHBH} and \ac{NSNS} populations are \acp{DCO} with more equal (i.e. $q\sim1$) mass ratio distributions compared to \ac{BHNS} (see also \S\ref{sec:Detectable-mass-distribution-functions}). The binaries that form merging \acp{BHBH} and \acp{NSNS} originate thus from different mass ratio populations at the same evolutionary stages compared to merging \acp{BHNS}.  This results in the $\beta$ parameter impacting the populations differently. Another difference is that the merging \ac{BHNS} typically form from different formation channels within isolated binary evolution compared to \ac{BHBH} and \ac{NSNS}. For example, in our fiducial model (A) a substantial fraction ($\gtrsim 25\%$ of all merging \acp{BHBH} in our simulation) of the \acp{BHBH} form through only stable mass transfer episodes, without engaging a \ac{CE},  whereas for \acp{BHNS} this is much smaller ($\lesssim 5\%$).  The parameter $\beta$ has a different impact on each of these formation channels. 

Second, the \ac{BHNS} rate increases with increasing values for $\alpha$ (models G, H, I and J) as a result of more efficient ejection of the \ac{CE}. This causes typically fewer systems to merge during the \ac{CE} \citep[cf.,][Figure 17]{Kruckow:2018}.  On the other hand, the rates slightly decline for the highest $\alpha_{\rm{CE}}$ values as in this case binaries do not shrink enough to merge in a Hubble time (see also the discussion on $\alpha$ in \S3.1.1 from \citealt{Bavera:2021}).

\subsection{GW detectable merger rates}
\label{subsec:comparing-BBH-BNS-detected-rates}

\begin{figure*}
    \centering
\includegraphics[width=1.0\textwidth]{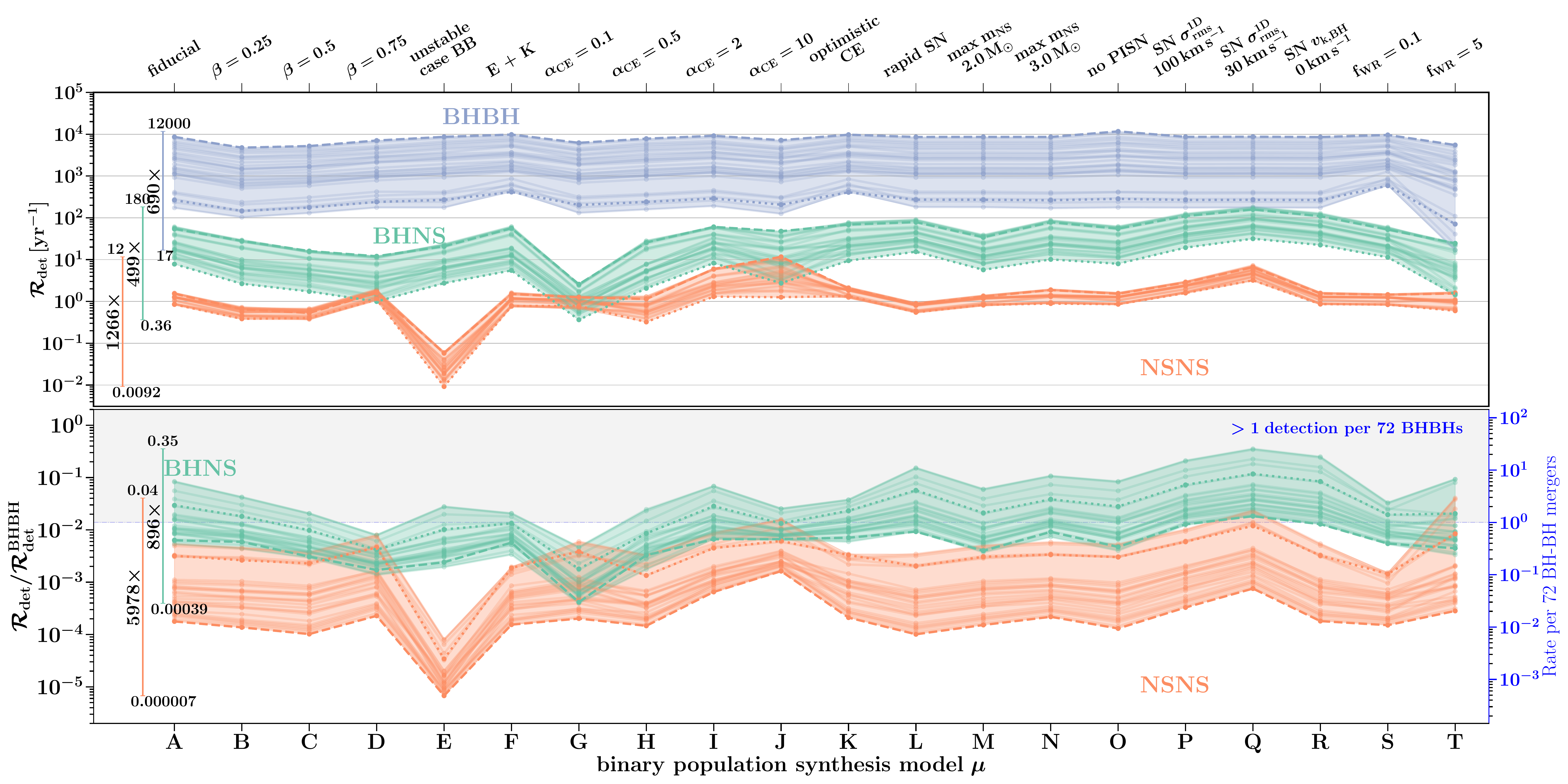} 
    \caption{ \textbf{Top panel:} The calculated detectable merger rates for merging BHBH (blue), BHNS (green) and NSNS (orange)  for a \ac{GW} network at design sensitivity for our \Nmodels  models.  \textbf{Bottom panel:} The predicted merging BHNS and NSNS detectable rate ratios over merging BHBH. On the right axes we scale the ratios to the 72 BHBH detections with false-alarm rate $\lesssim 1 \yearmin$ from the GWTC-3  \citep{Abbott:2021GWTC3pop}.      \textbf{Both panels:} The layout, lines and symbols are as in Figure~\ref{fig:IntrinsicRatesBBHBNSBHNS}. We note that the relative rates could be slightly, but not significantly, different at O1-O2-O3 (GWTC-3) sensitivity compared to design sensitivity. \href{https://github.com/FloorBroekgaarden/Double-Compact-Object-Mergers/blob/main/plottingCode/Fig_3/Rates_observed_ratio_bckground.pdf}{\faFileImage} \href{https://github.com/FloorBroekgaarden/Double-Compact-Object-Mergers/blob/main/plottingCode/Fig_3/make_figure_3.ipynb}{\faBook}}
    \label{fig:ObservedRatesRatiosBBHBNSBHNS}
\end{figure*}

The top panel in Figure~\ref{fig:ObservedRatesRatiosBBHBNSBHNS} shows the predicted detectable merger rates for our \Nmodels model variations.  These are calculated by volume integrating Equation~\ref{eq:MSSFR-merger-rate} and taking into account the sensitivity of a \ac{GW} detector, quantified by the probability \Pdet of observing a merging binary of specified masses at a given redshift and corresponding distance. Here we assume a detector sensitivity comparable to Advanced LIGO in its design configuration (hereafter, design sensitivity; \citealt{2015CQGra..32g4001L,2020LRR....21....3A}).  
We follow \citet{2018MNRAS.477.4685B} and choose a detector signal-to-noise ratio threshold of 8 as a proxy for detectability by a \ac{GW} detector network (e.g., LVK). The detectable merger rate is given by: 
\begin{align}
\label{eq:rate_detector}
	&\rate_{\rm{det}}(\tdet, m_{\rm{1}}, m_{\rm{2}}) = 
	\frac{\diff^3 \Ndet}{\diff \tdet  \diff m_{\rm{1}} \diff m_{\rm{2}}} =
	\notag  \\
	&\int 
	\diff \Vc  \,
	 \frac{\diff \tmerger}{\diff \tdet}  \,  
	  \frac{\diff^4 \Nmerger }{\diff \tmerger \diff \Vc   \diff m_{\rm{1}} \diff m_{\rm{2}}} \,
	 %
	  \Pdet (m_{\rm{1}}, m_{\rm{2}}, z),
\end{align}
where \tdet is the time in the detector (i.e., the observer) frame, and $m_{\rm{1}}$ and $m_{\rm{2}}$ are the component masses of the \ac{DCO} in the source frame (see  \citetalias{Broekgaarden:2021} for further details).  

Considering the \Nmodels model variations we find predicted detected rates for an LVK detector network at design sensitivity in the range $\rate_{\rm{det}}^{\rm{BHBH}} \approx  \RateObservedAzeroBHBHmin{-}\RateObservedAzeroBHBHmax\yearmin$, $\rate_{\rm{det}}^{\rm{BHNS}}\approx  \RateObservedAzeroBHNSmin{-}\RateObservedAzeroBHNSmax  \yearmin$ and $\rate_{\rm{det}}^{\rm{NSNS}}\approx \RateObservedAzeroNSNSmin{-}\RateObservedAzeroNSNSmax  \yearmin$, as shown in the top panel of Figure~\ref{fig:ObservedRatesRatiosBBHBNSBHNS}. We find that the stellar, binary, and cosmic evolution combined impact the predicted detectable merger rates by factors of up to $\approx 500{-}1300$. This is slightly higher compared to the intrinsic rates as the detectable population is biased to higher masses, where our simulations are relatively more sensitive to stellar evolution and \SFRD assumptions. 

\subsubsection{Relative merger rates}
The bottom panel of  Figure~\ref{fig:ObservedRatesRatiosBBHBNSBHNS} shows the relative merger rates between the different \ac{DCO} channels for LVK at design sensitivity. We find that almost all of our models predict a higher \ac{BHNS} detection rate compared to the \ac{NSNS} detection rate, except for model D (which has a mass transfer efficiency of $\beta=0.75$) and model G (which assumes  $\alpha_{\rm{CE}}=0.1$) in which a subset of the \SFRD models lead to higher \ac{NSNS} detection rates.  The higher \ac{BHNS} rate is a result from both the high intrinsic yield of \ac{BHNS} mergers compared to \ac{NSNS} mergers (Figure~\ref{fig:FormationRateDCO-per-metallicity}) and the larger detection volume for \ac{BHNS} compared to \ac{NSNS} mergers as a result from their larger masses (\S\ref{sec:Detectable-mass-distribution-functions}).

\subsection{GW detectable mass distribution functions}
\label{sec:Detectable-mass-distribution-functions}

\begin{figure*}
    \centering
\includegraphics[width=1\textwidth]{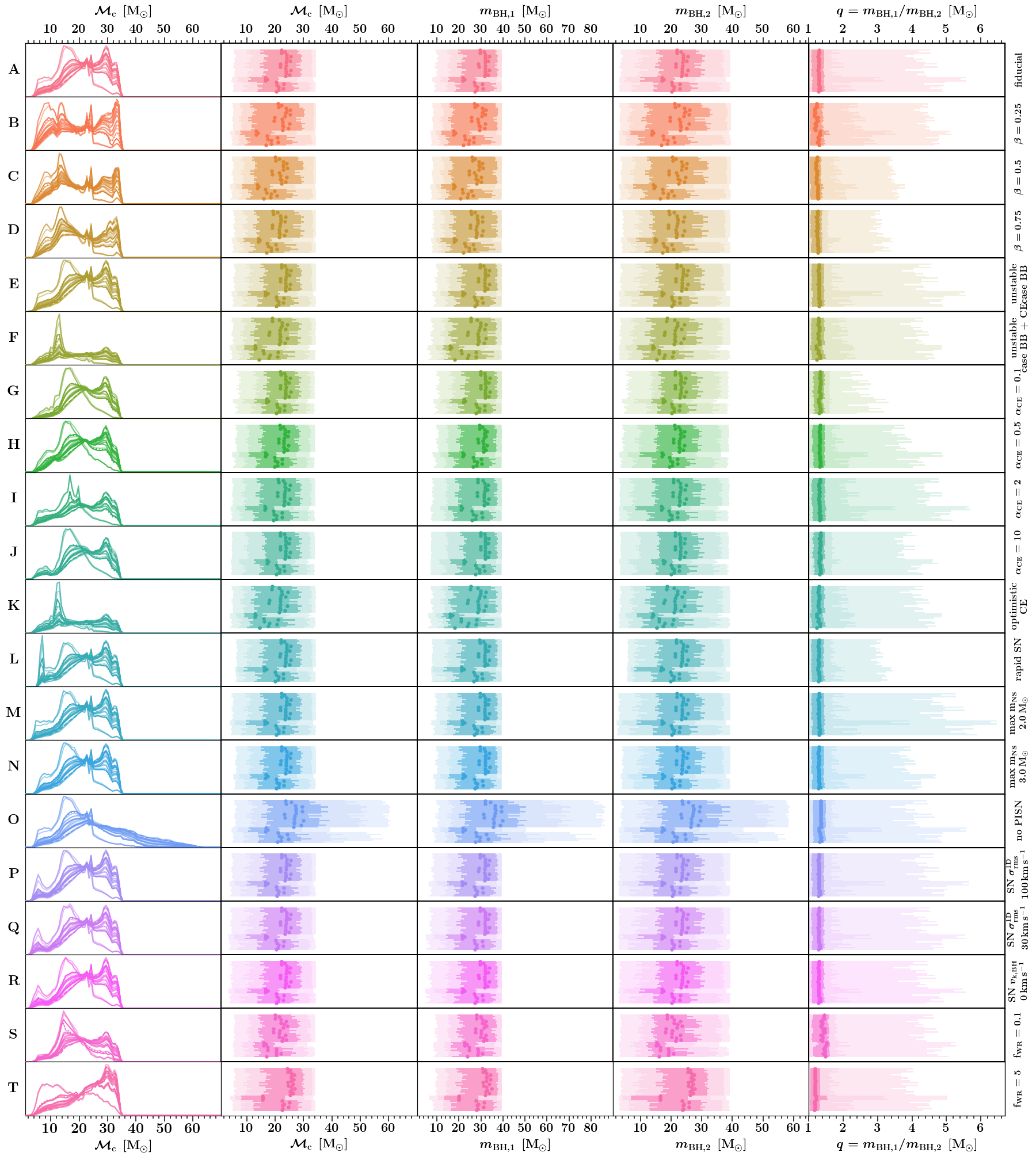} %
\caption{Shape of the detectable mass distribution functions for  BHBH mergers from our  \Nmodels model realizations. \textbf{First column:}  the probability distribution functions of the chirp mass  $\mchirpf$ at the time of merger. The distributions are normalized within each panel and given by $\diff \rate_{\rm{det}}/\diff x = (\diff^{2} \Ndet / \diff \tdet \diff x)$ for parameter $x$ (A variation of Equation~\ref{eq:rate_detector}). 
\textbf{Other columns:}  From left to right we show the chirp mass, the primary (most massive) and secondary (least massive) BH mass, and the mass ratio. In each sub-panel we show \NmodelsMSSFR  individual horizontal bars that visualize the median (scatter points) and the  $50\%$, $90\%$  and $99\%$  (three shades) distribution intervals.   A zoom-in that shows the order of the \SFRD models is given in Figure~\ref{fig:chirp-mass-percentile-zoom-in}. 
\textbf{Rows:} From top to bottom, the  panels/colors show the \NmodelsBPS different stellar evolution simulations from Table~\ref{tab:variations-BPS}.  Within each sub-panel we show the \NmodelsMSSFR \SFRD models with different lines (first column)  or horizontal bars (other columns).      \href{https://github.com/FloorBroekgaarden/Double-Compact-Object-Mergers/blob/main/plottingCode/Fig_4_and_Fig_5_and_Fig_6/KDEplot_massesdet_BBH.png}{\faFileImage} \href{https://github.com/FloorBroekgaarden/Double-Compact-Object-Mergers/blob/main/plottingCode/Fig_4_and_Fig_5_and_Fig_6/make_figure_4_5_and_6-Only_Masses.ipynb}{\faBook} }
    \label{fig:KDE-distributions-BHBH-masses}
\end{figure*}

\begin{figure*}
    \centering
\includegraphics[width=1\textwidth]{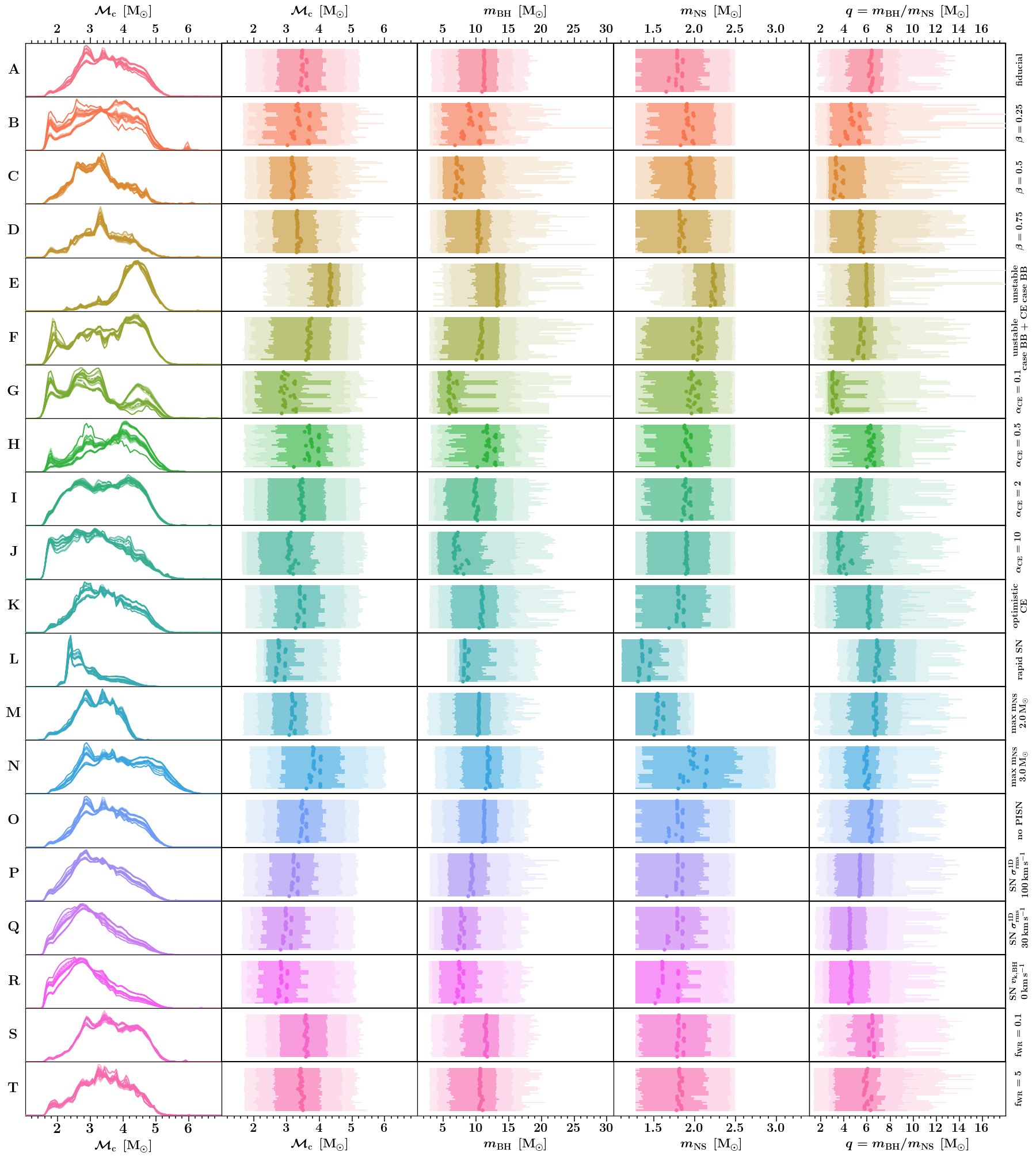} %
\caption{Same as Figure~\ref{fig:KDE-distributions-BHBH-masses} for detectable \ac{BHNS} mergers. For the individual masses we show the BH ($m_{\rm{BH}}$) and NS ($m_{\rm{NS}}$) masses.
\href{https://github.com/FloorBroekgaarden/Double-Compact-Object-Mergers/blob/main/plottingCode/Fig_4_and_Fig_5_and_Fig_6/KDEplot_massesdet_BHNS.png}{\faFileImage} \href{https://github.com/FloorBroekgaarden/Double-Compact-Object-Mergers/blob/main/plottingCode/Fig_4_and_Fig_5_and_Fig_6/make_figure_4_5_and_6-Only_Masses.ipynb}{\faBook} 
}
    \label{fig:KDE-distributions-BHNS-masses}
\end{figure*}

\begin{figure*}
    \centering
\includegraphics[width=1\textwidth]{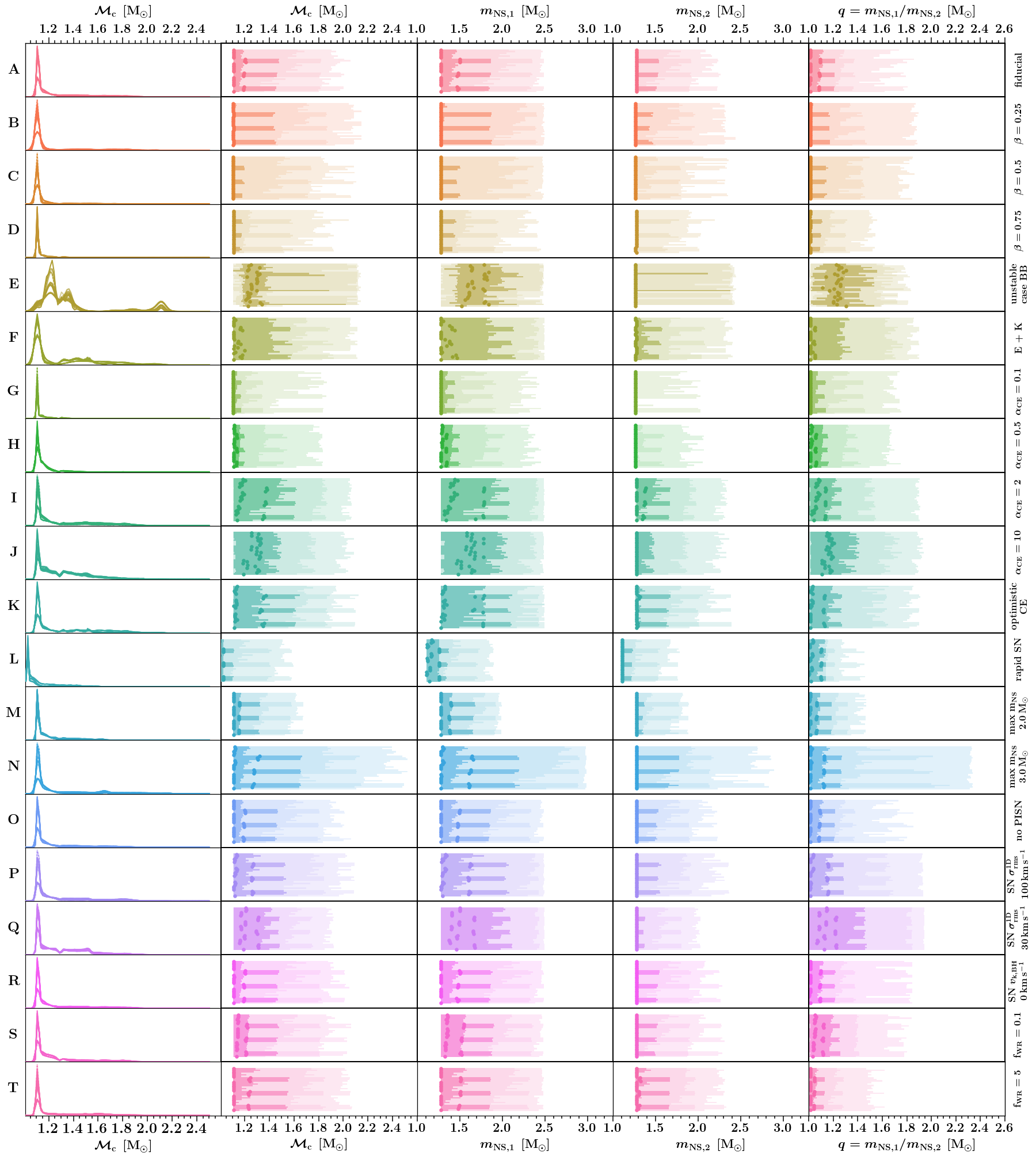} %
\caption{Same as Figure~\ref{fig:KDE-distributions-BHBH-masses} for detectable \ac{NSNS} mergers. 
For the individual masses we show the primary (most massive, $m_{\rm{NS,1}}$) and secondary (least massive,  $m_{\rm{NS,2}}$) NS mass. 
The shape of model E (unstable case BB) suffers significantly from sampling noise due to the low number of NSNS systems in this variation. 
\href{https://github.com/FloorBroekgaarden/Double-Compact-Object-Mergers/blob/main/plottingCode/Fig_4_and_Fig_5_and_Fig_6/KDEplot_massesdet_BNS.png}{\faFileImage} \href{https://github.com/FloorBroekgaarden/Double-Compact-Object-Mergers/blob/main/plottingCode/Fig_4_and_Fig_5_and_Fig_6/make_figure_4_5_and_6-Only_Masses.ipynb}{\faBook} }
    \label{fig:KDE-distributions-NSNS-masses}
\end{figure*}

The uncertainties in stellar and binary evolution and \SFRD also impact the shapes of the mass distributions of detectable \ac{DCO} mergers, in addition to the overall merger rate densities. We show this in  Figures~\ref{fig:KDE-distributions-BHBH-masses} (BHBH), \ref{fig:KDE-distributions-BHNS-masses} (BHNS) and \ref{fig:KDE-distributions-NSNS-masses} (NSNS) for our \Nmodels model realizations. Each Figure shows the normalized distributions for the \ac{DCO} component masses $m_1$ and $m_2$, chirp mass $\mchirpf = {(m_{\rm{1}} m_{\rm{2}})^{3/5}}/{(m_{\rm{1}} + m_{\rm{2}})^{1/5}}$, and mass ratio $q = m_1 / m_2$, where we use subscripts `1' and `2' to indicate the more massive and less massive component in the double compact object system, respectively.  
To compare the shapes we show kernel density distributions for the chirp mass and summary statistics (i.e., median, 50, $90$ and $99$ distribution percentiles) for the individual masses, chirp mass and mass ratio for the \ac{BHBH}, \ac{BHNS} and \ac{NSNS} mergers. All distributions are weighted for the detection volume and sensitivity of LVK at design sensitivity using Equation~\ref{eq:rate_detector} and given by the differential detectable merger rate $\diff^{2} \Ndet / \diff \tdet \diff x$, with $x$ being one of the mass parameters mentioned above.\footnote{From hereon we will use the short notation $\diff \mathcal{R}_{\rm{det}} / \diff x$ for this differential detection rate that describes the `shape' of the distributions.}


\subsubsection{Impact from binary star and cosmic evolution on the distribution shapes}
\label{sec:Impact-from-massive-binary-star-and-cosmic-evolution-on-the-distribution-shapes}
We qualitatively compare the impact on the shape of the detectable mass distributions from variations in stellar and binary evolution and variations in \SFRD by analyzing how the distributions (and distribution statistics) vary; see Figures~\ref{fig:KDE-distributions-BHBH-masses}, \ref{fig:KDE-distributions-BHNS-masses} and \ref{fig:KDE-distributions-NSNS-masses}. Distributions that are similar between vertical panels (different colors) indicate that changes in stellar and binary evolution assumptions  do not significantly impact the distribution shape. On the other hand,  distributions that are similar between variations plotted within one sub-panel indicate these \SFRD realizations do not significantly impact the shape of the distributions. By comparing these two effects directly we can qualitatively analyze which of the two uncertainties dominates the shape of the detectable mass distributions.

For merging BHBH systems we find that the uncertainty in the shape of the mass distribution is significantly impacted by both variations in stellar evolution and the \SFRD. The impact from the \SFRD reflects the metallicity dependence of BHBH formation (\S\ref{section:Formation-rates-per-metallicity}), which also strongly impacts the resulting \ac{BH} masses \citep[cf.,][]{Belczynski:2010,GiacobboMapelli:2018}. For example, the peak in the detectable \ac{BHBH} chirp mass distribution around $30$--$35$\Msun that is visible in the majority of models in Figure~\ref{fig:KDE-distributions-BHBH-masses} disappears in a small subset of the \SFRD models. These are the \SFRD models with the \citet{2004MNRAS.355..764P} \ac{GSMF} and  \citet{2016MNRAS.456.2140M}  \ac{MZR} ($\rm{xyz} = $113, 213 and 313), which correspond to realizations with high average metallicities compared to the other \SFRD models (see Figure~B1 in  \citetalias{Broekgaarden:2021}) and lead to fewer massive \ac{BHBH} mergers. The stellar and binary evolution variation that impacts the mass distributions most drastically is the model in which we assume pair-instability \acp{SN} do not occur (model O), leading to the formation of \acp{BH} with masses $\gtrsim 40\Msun$. This model assumption may be unrealistic \citep[e.g.,][]{2017ApJ...836..244W, Farmer:2019}. Other significant changes, best visible in the kernel density functions (left-most column in Figure~\ref{fig:KDE-distributions-BHBH-masses}), are present for models K (optimistic CE), L (rapid SN remnant model) and T ($f_{\rm{WR}}=0.1$), where, for example, the chirp mass peaks shift,  disappear or are created compared to the fiducial model A (e.g., in the `optimistic CE' model due to the many additional  \ac{BHBH} systems added around lower chirp masses). 

For merging \ac{BHNS} systems, on the other hand, the  distribution shapes are predominantly impacted by the stellar and binary evolution model variations compared to the \SFRD variations (Figure~\ref{fig:KDE-distributions-BHNS-masses}). Primary effects include models E (unstable case BB), L (rapid SN), M (max $\mnsf =2\Msun$) and N (max $\mnsf =3\Msun$), where the median and distribution percentiles are visibly  different compared to the fiducial model A. 
 Our \SFRD variations do not significantly impact the distribution shapes (i.e. changing $\diff \mathcal{R}_{\rm{det}} / \diff x$ more than a factor 2) for most of the models except model B, G and H.

For NSNS systems the shape of the normalized mass distributions are impacted by both sets of variations (Figure~\ref{fig:KDE-distributions-NSNS-masses}).  Among the binary stellar evolution variations, the models that impact the shape of the distribution significantly are realizations that change the SN remnant mass or kick velocity, and/or the CE efficiency (models G, H, I, J, K, L,  M, N, P and Q). The distributions from model E (unstable case BB mass transfer) are dominated by sampling noise as a result of the low NSNS yield in this model (\S\ref{section:Formation-rates-per-metallicity}). For the \SFRD particularly all our models with the  \citet{2006ApJ...638L..63L} \ac{MZR} model ($\rm{z}=1$) significantly impact the \ac{NSNS} mass distributions, shifting the distributions to higher median masses in the panels of Figure~\ref{fig:KDE-distributions-NSNS-masses}.

In summary, we find that for \acp{BHNS}, the variations in the distribution shapes are typically dominated by stellar and binary evolution assumptions, suggesting that observations of \ac{BHNS} systems could aid in constraining stellar evolution models.  For merging \ac{BHBH} and \ac{NSNS} the distribution shapes are impacted by both sets of variations (Figures~\ref{fig:KDE-distributions-BHBH-masses} and \ref{fig:KDE-distributions-NSNS-masses}). We therefore argue that constraining stellar evolution or \SFRD models solely from the distribution shapes of merging BHBHs or NSNSs may be challenging. We discuss this further in \S\ref{sec:discussion}.

\subsubsection{Distribution properties considering all model realizations}
\label{sec:results-distributions-BH-mass}

In the previous section we showed that the model uncertainties in binary and stellar evolution, and \SFRD can significantly impact the shapes of the \ac{DCO} mass distributions. Here we instead discuss several specific examples of features in the mass distributions in  Figures~\ref{fig:KDE-distributions-BHBH-masses}, \ref{fig:KDE-distributions-BHNS-masses} and \ref{fig:KDE-distributions-NSNS-masses} that are robust across all \Nmodels model variations explored in this work.

\begin{itemize}
\item First, in all \Nmodels model variations $95\%$ ($99\%$) of the detectable \ac{BHBH} mergers have mass ratios $\qf\lesssim 4$ ($\qf\lesssim 6$) (right-most column in Figure~\ref{fig:KDE-distributions-BHBH-masses}). The \ac{BHBH} distribution medians and percentiles furthermore indicate that in all model variations \acp{BHBH} from isolated binary evolution prefer order unity mass ratios. Detecting a significant fraction ($\gtrsim5\%$) of \ac{BHBH} mergers with large mass ratios ($\qf\gtrsim5$) would point to other formation pathways or missing physics in our simulations. This is consistent with the 72 \ac{BHBH} detections announced by LVK with a false alarm rate $ < 1 \yearmin$, which most of are inferred to have mass ratios consistent with unity \citep{Abbott:2021GWTC3}. However, the \ac{BHBH} merger GW190412 ($1/\qf = 0.28_{-0.07}^{+0.12}$; \citealt{Abbott:2020gw190412}) and the \ac{BHBH} merger candidate GW190814 ($1/\qf = 0.112_{-0.009}^{+0.008}$; \citealt{Abbott:2020gw190814}), if common, could hint to the existence of a population with more extreme mass ratios \citep[see also][and references therein]{ArcaSedda:2021,Lu:2021,Zevin:2020-lower-mass-gap}.

\item Second, $\gtrsim 95\%$ of all \ac{BHBH} mergers are expected to contain a \ac{BH} with $\gtrsim 8\Msun$ in all \Nmodels model variations (third column of Figure~\ref{fig:KDE-distributions-BHBH-masses}). This is not the case for \acp{BH} in \ac{BHNS} systems where typically $\gtrsim 25\%$ of the detectable mergers are expected to contain a \ac{BH} of $\lesssim 8\Msun$ (Figure~\ref{fig:KDE-distributions-BHNS-masses}). Our models predict the secondary \ac{BH} in the population of detected \ac{BHBH} mergers to also commonly be massive ($\gtrsim 8\Msun$), although many of our models do allow for at least $\gtrsim 25\%$ of the  \ac{BHBH} mergers to contain a secondary BH with $\lesssim 10\Msun$. For the currently reported \ac{BHBH} detections by LVK with a false alarm rate $ < 1 \yearmin$ (but even for those with a false alarm rate $ < 0.25\yearmin$) all of the inferred medians of the primary \ac{BH} mass have \ac{BH} masses $\mbhfone \gtrsim 8\Msun$ \citep[][Table I]{Abbott:2021GWTC3pop}, consistent with our models.

\item Third, in all \Nmodels model variations $\lesssim 5\%$ of the detectable merging \ac{BHNS} systems have \ac{BH} masses $\mbhf\gtrsim 18\Msun$ (third column of Figure~\ref{fig:KDE-distributions-BHNS-masses}). As discussed in  \citetalias{Broekgaarden:2021}, this is due to the fact that more equal mass stars more readily survive important stellar evolution phases in the formation pathways to \ac{BHNS} mergers; to form a \ac{BHNS} one of the stars needs to be of sufficiently low mass to form a \ac{NS}, leading to a preference for the other star to also be of lower mass than is typical in \ac{BHBH}.  Consequently, we find that in all our \Nmodels model realizations $\lesssim 5\%$ of \ac{BHNS} mergers have chirp masses of $\gtrsim 5.5 \Msun$.  This suggests that detecting a \ac{BHNS} merger with a \ac{BH} mass  $\mbhf\gtrsim 18\Msun$ could indicate that the system did not form from isolated binary evolution processes our models include, but instead from chemically homogeneous evolution and/or dynamical formation where such high \ac{BH} masses are more commonly expected \citep[e.g.,][]{2017A&A...604A..55M,2020MNRAS.498.4088M,2020MNRAS.497.1563R}.  The two detected \ac{BHNS} systems with a false alarm rate $< 1\yearmin$, GW200105 and GW200115, have $\mchirpf\lesssim 3.5 \Msun$ (at the $90\%$ credible interval) \citep[][]{Abbott:2021-first-NSBH} consistent with our models \citep[see also][]{BroekgaardenBerger2021,Broekgaarden:2021}. 

\item Fourth, we find that in all \Nmodels model variations the \acp{NS} in detectable \ac{NSNS} mergers typically have lower \ac{NS} masses compared to those in detectable \ac{BHNS} mergers  (Figures \ref{fig:KDE-distributions-BHNS-masses} and \ref{fig:KDE-distributions-NSNS-masses}). We find this is true for both NS components in the \ac{NSNS} mergers. This is again due to the preferrence for equal mass binaries to survive important evolutionary phases leading to \ac{BHNS} formation, thereby favoring more massive pre-\ac{NS} stars to form a \ac{BHNS} system (see  \citetalias{Broekgaarden:2021} for more details). To date two  \ac{NSNS} and \ac{BHNS} detections have been announced by LVK with a false alarm rate $ < 1 \yearmin$ \citep{Abbott:2021GWTC3pop}. For the \ac{NSNS} detections the \ac{NS} masses are $\mnsfone=1.46_{-0.10}^{+0.12}$ and $\mnsftwo = {1.27}_{-0.09}^{+0.09}$  (GW170817; \citealt{Abbott:17gw170817discovery}) and $\mnsfone=2.0_{-0.3}^{+0.6}$ and $\mnsftwo = {1.4}_{-0.3}^{+0.3}$ (GW190425; \citealt{Abbott:2020gw190425}), which may indicate a \ac{NSNS} mass distribution that is not consistent with Galactic \ac{NSNS} observations \citep[e.g.][]{VignaGomez:2018, Abbott:2021GWTC3pop}. For the \ac{BHNS} detections the \ac{NS} masses were found to be $\mnsf = 1.{9}_{-0.2}^{+0.3}$ $\Msun$ (GW200105) and $\mnsf = 1.{5}_{-0.3}^{+0.7}$\Msun (GW200115) \citep{Abbott:2021-first-NSBH}. More detections are needed to calculate robust median \ac{NS} masses in \ac{NSNS} and \ac{BHNS} detections. 
\end{itemize}

Besides the four specific points above, more common trends throughout our \Nmodels model variations are visible in Figures~\ref{fig:KDE-distributions-BHBH-masses}, \ref{fig:KDE-distributions-BHNS-masses} and \ref{fig:KDE-distributions-NSNS-masses}, particularly in the lower and upper bounds for the \ac{BH} and \ac{NS} masses. Several of these trends, however, are direct results from the remnant mass prescriptions used in our models. For example, all models with the exception of model O (no PISN) predict that $90\%$ of the detected \ac{BHBH} mergers have chirp masses in the range $\mchirpf \approx 5$--$35\Msun$. The upper limit is set by the maximum possible BH masses of $\sim 40\Msun$ allowed due to the implementation of pair-instability \ac{SN} in our models \citep[e.g.,][]{Marchant:2019,Stevenson:2019}; in model O we do not implement pair-instability \acp{SN}, and find \ac{BH} masses up to about $80\Msun$ and \ac{BHBH} chirp masses up to about $60\Msun$. 
In addition, in most of our model realizations the majority of \acp{NS} in \ac{NSNS} mergers have masses of $\approx 1.3$\Msun (consistent with e.g., \citealt{2017ApJ...846..170T,VignaGomez:2018}). This is because many \ac{NSNS} experience at least one \ac{ECSN}, which in \COMPAS are mapped to a mass of $1.26\Msun$ (\citealt{1996ApJ...457..834T}, \citealt{COMPAS:2021methodsPaper}).  In addition, many \ac{NSNS} mergers in our models originate from relatively low mass stars with carbon-oxygen cores of $\lesssim 2.5\Msun$ at the time of the \ac{SN}, which in the delayed \ac{SN} remnant mass prescription are mapped to a fixed value of $\mnsf \approx 1.28\Msun$  \citep{2012ApJ...749...91F}. Combined, this results in a peak of \acp{NS} masses at $\mnsf\approx 1.3$\Msun. Exceptions to this are the rapid \ac{SN} remnant mass model (model I) where these low mass stars are instead mapped to \ac{NS} masses of $\approx 1.1\Msun$ leading to a broader peak at $\approx 1.1-1.3\Msun$. The lower and upper \ac{NS} limits are also artificially set in our models to $\approx 1\Msun$ and $\approx 2.5\Msun$, respectively \citep[cf.,][]{2012ApJ...749...91F}, except in models M and N where we set the upper limit to 2\Msun and 3\Msun, respectively.


\subsection{Delay time and birth metallicity distributions}
\label{sec:results-distributions-delay-times}

\begin{figure*}
    \centering
\includegraphics[width=1\textwidth]{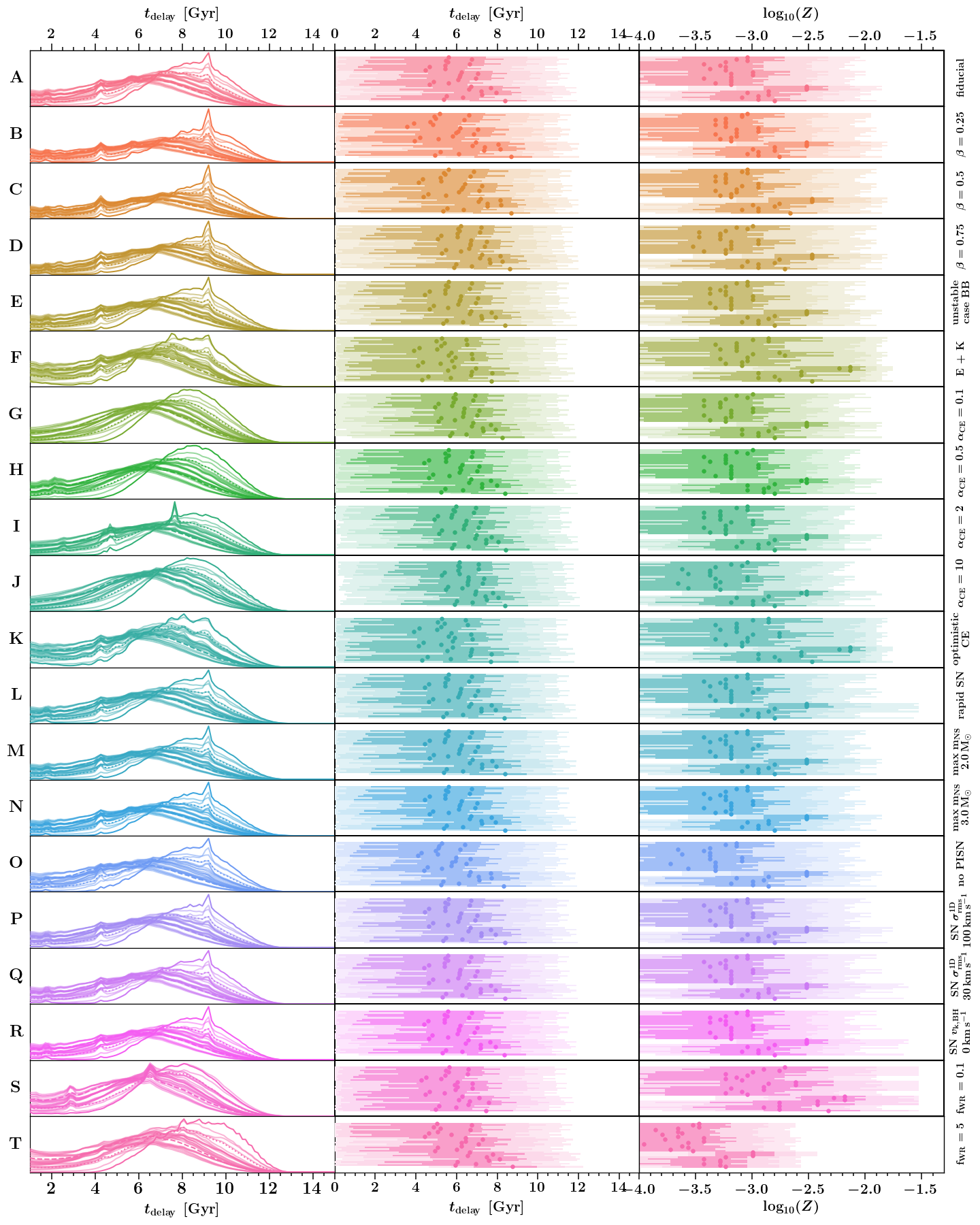} %
\caption{The delay time  $\tdelay$ and birth metallicity \Zi distributions for detectable BHBH mergers for our \Nmodels model realizations. The distributions and percentiles are plotted as in Figure~\ref{fig:KDE-distributions-BHBH-masses}. The delay time is the time since formation of the binary system at the zero-age main sequence until the moment of merger. \href{https://github.com/FloorBroekgaarden/Double-Compact-Object-Mergers/blob/main/plottingCode/Fig_7_and_Fig_8_and_Fig_9/KDEplot_time_metallicity_det_BBH.png}{\faFileImage} \href{https://github.com/FloorBroekgaarden/Double-Compact-Object-Mergers/blob/main/plottingCode/Fig_7_and_Fig_8_and_Fig_9/make_Fig_7_and_Fig_8_and_Fig_9.ipynb}{\faBook}}
    \label{fig:KDE-distributions-BHBH-delay-times}
\end{figure*}

\begin{figure*}
    \centering
\includegraphics[width=1\textwidth]{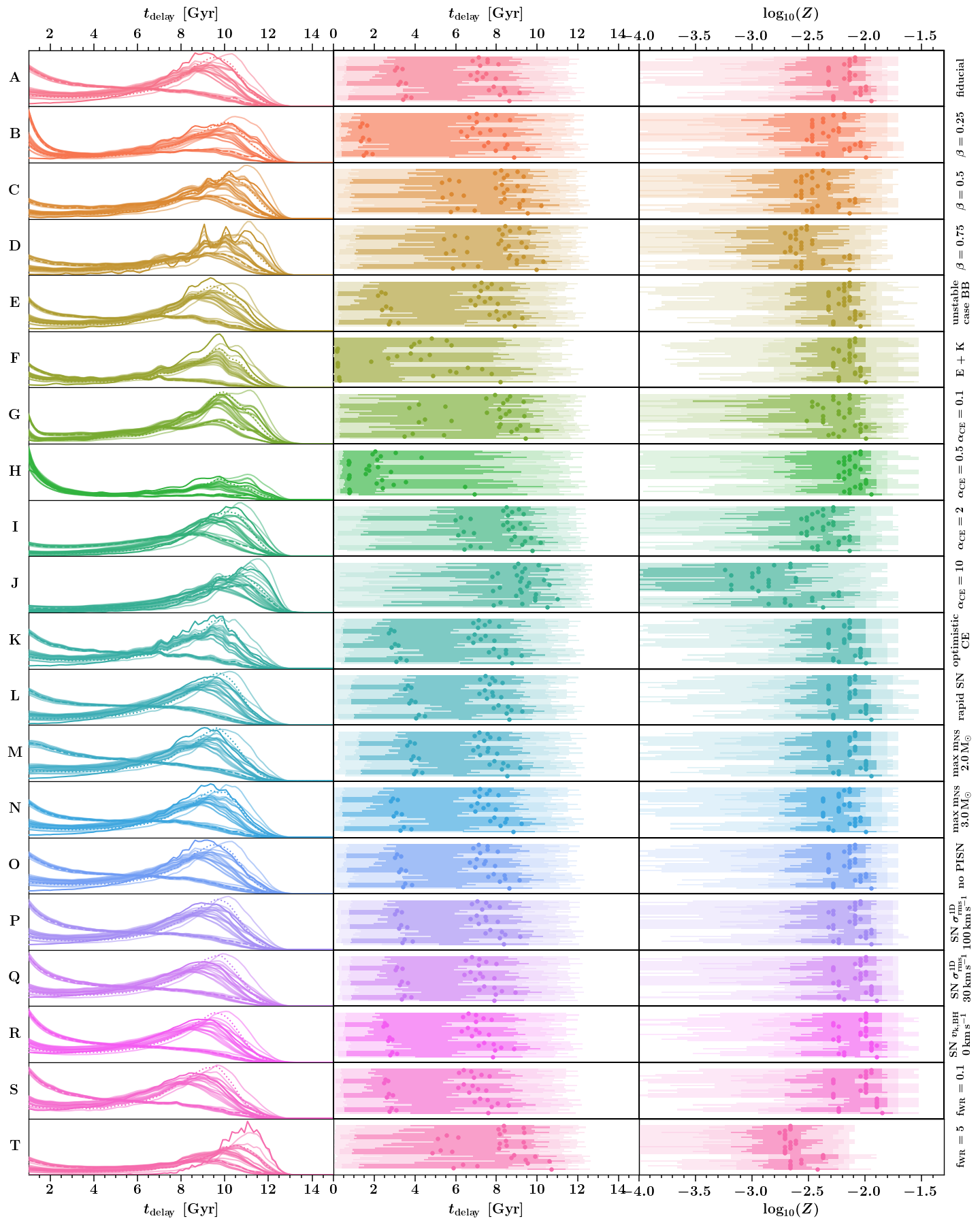} %
\caption{Same as Figure~\ref{fig:KDE-distributions-BHBH-delay-times} for detectable \ac{BHNS} mergers. \href{https://github.com/FloorBroekgaarden/Double-Compact-Object-Mergers/blob/main/plottingCode/Fig_7_and_Fig_8_and_Fig_9/KDEplot_time_metallicity_det_BHNS.png}{\faFileImage} \href{https://github.com/FloorBroekgaarden/Double-Compact-Object-Mergers/blob/main/plottingCode/Fig_7_and_Fig_8_and_Fig_9/make_Fig_7_and_Fig_8_and_Fig_9.ipynb}{\faBook}}
    \label{fig:KDE-distributions-BHNS-delay-times}
\end{figure*}

\begin{figure*}
    \centering
\includegraphics[width=1\textwidth]{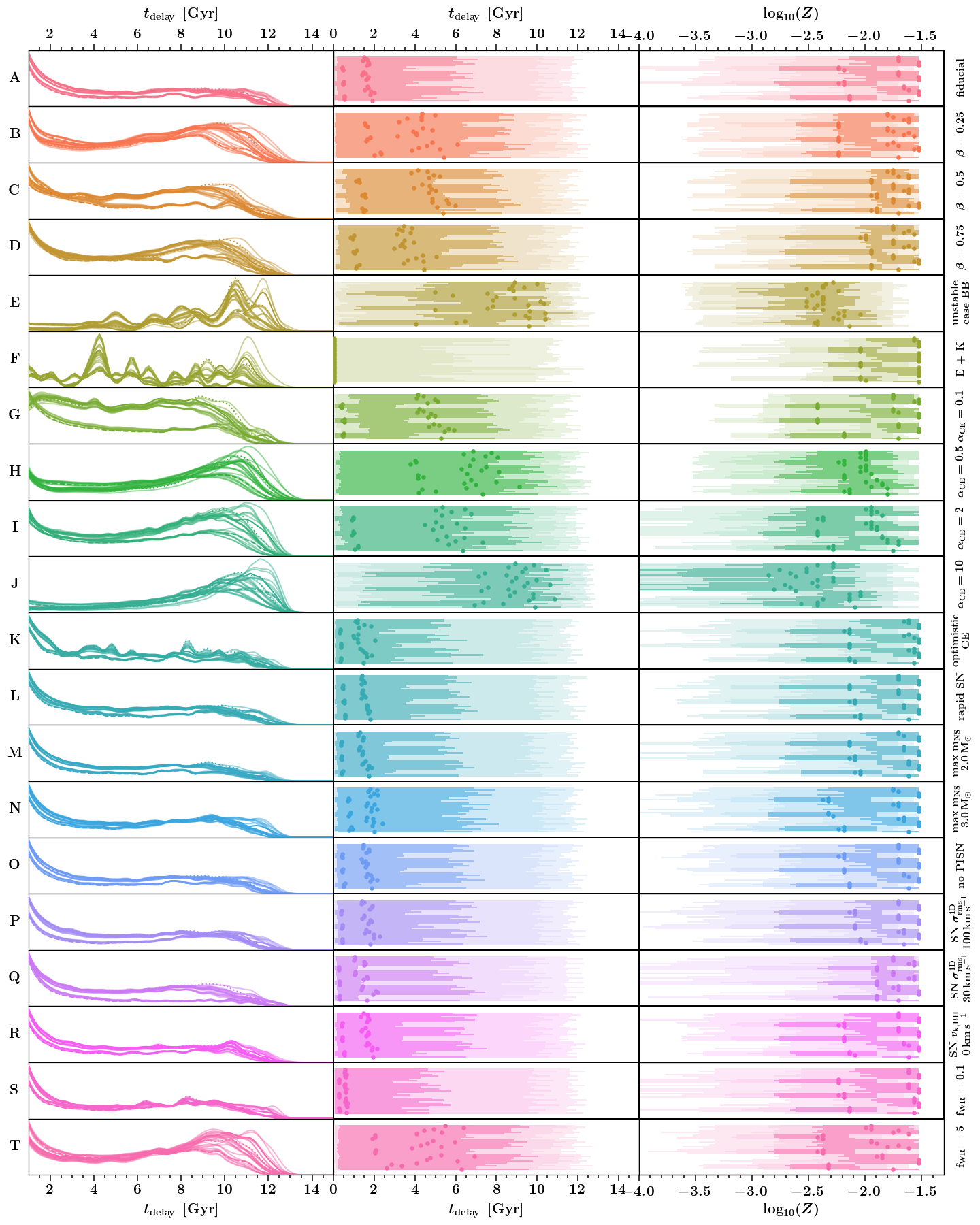} %
\caption{Same as Figure~\ref{fig:KDE-distributions-BHBH-delay-times} for detectable \ac{NSNS} mergers. The shape of model E (unstable case BB) suffers significantly from sampling noise due to the low number of NSNS systems in this variation (and similarly for model F). \href{https://github.com/FloorBroekgaarden/Double-Compact-Object-Mergers/blob/main/plottingCode/Fig_7_and_Fig_8_and_Fig_9/KDEplot_time_metallicity_det_BNS.png}{\faFileImage} \href{https://github.com/FloorBroekgaarden/Double-Compact-Object-Mergers/blob/main/plottingCode/Fig_7_and_Fig_8_and_Fig_9/make_Fig_7_and_Fig_8_and_Fig_9.ipynb}{\faBook}}
    \label{fig:KDE-distributions-NSNS-delay-times}
\end{figure*}

The uncertainties in the stellar evolution and \SFRD prescriptions in population synthesis modelling can also impact the expected delay time (\tdelay) and birth metallicity (\Zi) distributions of the detectable \ac{DCO} mergers. We show the impact from the \Nmodels model realizations explored in this study in Figures~\ref{fig:KDE-distributions-BHBH-delay-times}, ~\ref{fig:KDE-distributions-BHNS-delay-times} and ~\ref{fig:KDE-distributions-NSNS-delay-times}.  We focus on the \tdelay and \Zi distributions as these properties can be (indirectly) constrained from observations \citep[e.g.,][]{Im:2017,Safarzadeh:2019delayTimes,FishbachKalogera:2021delayTimes}. In addition, these distributions provide insights into the range of birth metallicities and star formation redshifts that current GW observations probe, which can be informative for future modelling (e.g., help understand where to focus computational time and which \Zi to simulate). 
We note that impacts on the \tdelay distribution also impact (indirectly) the distribution of \Zi, because the birth metallicity distribution ($\diff P/ \diff Z$; \S\ref{sec:method-metallicity-specific-star-formation-rate-density-prescription}) evolves as a function of redshift. 

The most striking feature of the \tdelay and \Zi distributions is that both are significantly impacted by the \SFRD variations for all three \ac{DCO} merger types. This is particularly discernible in the large scatter of the distribution percentiles (see each sub-panel). We find that the impact from \SFRD models is particularly significant for our variations in the choice of the metallicity probability function ($\diff P / \diff Z$), which describes the distribution of birth metallicities of stars at a given redshift. This is because the birth metallicity strongly impacts the properties and rate of the \ac{DCO} mergers, including through the metallicity-dependent formation yields, through mass loss that impact the \ac{DCO} masses and can widen the orbit of the binary, and through the metallicity-dependent radial expansion of stars, which impacts mass transfer phases in our models (\S\ref{section:Formation-rates-per-metallicity}). In the convolution to obtain the detectable \ac{DCO} mergers in Equation~\ref{eq:rate_detector} all of these properties impact the resulting population. Thus, the \tdelay and \Zi distributions of the detectable \ac{DCO} mergers are significantly impacted by the choice of ($\diff P / \diff Z$) and \SFRD. 

Variations in the stellar and binary evolution assumptions can also significantly impact the \tdelay and \Zi distributions. For example, model T ($f_{\rm{WR}}=5$), which increases the mass loss through stellar winds, drastically suppresses the number of \ac{BHBH} and \ac{BHNS} events that can form at higher metallicities ($\log(\Zi)\gtrsim -2.5$), thereby leading to fewer of the detectable \ac{BHBH} and \ac{BHNS} mergers forming from these metallicities (Figures~\ref{fig:KDE-distributions-BHBH-delay-times} and \ref{fig:KDE-distributions-BHNS-delay-times}). 
Another example is Model J ($\alpha_{\rm{CE}}=10$), which results in longer \tdelay and lower \Zi compared to our fiducial model in Figures~\ref{fig:KDE-distributions-BHNS-delay-times} and ~\ref{fig:KDE-distributions-NSNS-delay-times}. 
In model J orbital angular momentum of the binary can much more efficiently be transformed into ejecting the \ac{CE}. 
This leads to less orbital shrinking during the \ac{CE} phase and longer inspiral times. 
As a result only \acp{BHNS} and \acp{NSNS} that formed early in the Universe with low \Zi have had long enough the time to inspiral and be detectable today. 
Model J does not impact the \ac{BHBH} distributions as significantly because most detectable \ac{BHBH} mergers in our simulations only go through stable mass transfer phases (as these produce more massive \acp{BHBH} in our simulations that the detectable population is biased to, cf. \citealt[][]{vanSon:2021}).

Overall, we find that the delay time distributions span a broad range between a few Myr and the Hubble time with possible peaks both at short ($ < 1 \Gyr$) and long ($> 5 \Gyr$) delay times, reflecting contributions from binaries born at lower and higher redshifts, respectively. This is a result from the interplay between the \SFRD being a function of redshift in combination with the \ac{DCO} properties and formation efficiencies being \Zi dependent.   We find that our model variations indicate that the detectable \ac{BHBH} mergers (Figure~\ref{fig:KDE-distributions-BHBH-delay-times}) probe systems with the longest median delay times, compared to \ac{BHNS} and \ac{NSNS} mergers (Figures~\ref{fig:KDE-distributions-BHNS-delay-times} and \ref{fig:KDE-distributions-NSNS-delay-times}). However, exceptions exist, including model J ($\alpha_{\rm{CE}}=10$) where the median delay times of detected \ac{BHNS} and \ac{NSNS} mergers are larger than those of \ac{BHBH} mergers for the reasons discussed above.

For the birth metallicity distribution shapes the right panels of Figures~\ref{fig:KDE-distributions-BHBH-delay-times}, \ref{fig:KDE-distributions-BHNS-delay-times} and \ref{fig:KDE-distributions-NSNS-delay-times} show that the detectable \ac{DCO} mergers arise from a broad range of birth metallicities. As expected from the formation yield efficiencies, BHBH mergers typically originate from the lowest birth metallicities, with most of our models having a median $\log_{10}(\Zi)$ of $\lesssim -2.5$, whereas the detectable \ac{BHNS} and \ac{NSNS} populations originate in our models from higher \Zi with  most models having median $\log_{10}(\Zi)$ of $\gtrsim -2.5$.

%


\section{Discussion}
\label{sec:discussion}

\subsection{A realistic view of population synthesis models}
\label{sec:discussion-pop-synth-is-uncertain}

In this paper we demonstrate the importance of considering the uncertainties in both massive star and binary evolution, as well as the \SFRD when aiming to learn about the formation, evolution, and mergers of binaries through a comparison of population synthesis models to GW observations. Here we discuss the implications of our results for interpreting population synthesis studies. 

On the one hand, our findings provide a cautionary note for using GW observations to uncover stellar and binary evolution properties. 
 Studies that draw conclusions by comparing population synthesis results with \ac{GW} observations without considering the wide range of uncertainties in both stellar/binary evolution and \SFRD assumptions could be drastically biased by the specific model realization that is chosen to derive the results \citep[cf.][]{Belczynski:2021,Bouffanais:2021}. 
For example, studies including \citet{Zevin:2017}, \citet{Bouffanais:2019},  \citet{Franciolini:2021}, \citet{Ng:2021} and \citet{Zevin:2021-branching-ratios} aim to determine the contributions of different formation channels to the observed \ac{BHBH} population.
Further examples include \citet{Fragione:2021}, which estimates the number of \ac{BHNS} mergers with an electromagnetic counterpart.
Our work shows that uncertainties in both stellar and binary evolution and \SFRD modelling can drastically impact the rate and distribution shapes of the \ac{BHBH}, \ac{BHNS} and \ac{NSNS} populations, challenging the ability to draw strong conclusions when only considering a few population synthesis models at this stage.  

The situation is even more complex when considering that our larger suite of models (\Nmodels realizations) still only represents a subset of the overall uncertainties in population synthesis studies. For example, even our broader set of models does not account for uncertainties in the stellar evolution tracks \citep[e.g.,][]{2020A&A...637A...6L,2020MNRAS.497.4549A}, internal mixing \citep[e.g.,][]{Schootemeijer:2019}, stellar rotation \citep[e.g.,][]{deMinkMandel:2016,Mapelli:2020}, the more complex physics of the CE phase \citep[e.g.,][]{Klencki:2021, Ivanova:2020book,2021A&A...650A.107M, Olejak:2021CE}, the additional possible remnant mass prescriptions \citep[e.g.][]{Dabrowny:2021,Mandel:2021MNRAS}, the initial conditions of binary systems \citep[e.g.,][]{deMinkBelczynski:2015,2017ApJS..230...15M,2018A&A...619A..77K}, and the possible contributions from other formation channels \citep[e.g.,][]{Zevin:2021-branching-ratios}. In addition, there are \SFRD model variations that we did not explore, including alternative analytical prescriptions \citep[e.g.,][]{Chruslinska:2019dco,Tang:2020}, prescriptions derived from cosmological (zoom-in) simulations (e.g., \emph{FIRE}, \emph{Illustris}, \emph{EAGLE}, \emph{Millennium};  \citealt{Lamberts:2016,2017MNRAS.472.2422M, duBuisson2020,Briel:2021, Chu:2021}), and prescriptions inferred more directly from observations \citep[e.g., ][]{Chruslinska:2019obsSFRD,Chruslinska:2021}.

On the other hand, the fact that model predictions are sensitive to the model assumptions also has positive consequences. As the true \ac{DCO} merger rates and properties become more constrained by observations, then identifying which models are inconsistent with reality aids in excluding some combinations of model assumptions, and so helps to constrain stellar evolution and cosmic history. To do this we need to sufficiently understand how the uncertainties in our assumptions affect the model output.  
The results in this paper show that simultaneously modelling the impact of uncertainties from stellar evolution and \SFRD on the rates and distribution shapes of all three \ac{DCO} merger types can help.  For example, in \S\ref{sec:results-comparing-BBH-BNS-intrinsic-rates} we found that the \ac{BHBH} merger rate densities predicted from binary population synthesis are relatively sensitive to \SFRD uncertainties and might thereby present a good test bed to constrain \SFRD prescriptions. Similarly, we found that the \ac{NSNS} merger rate densities might be a good test bed to constrain stellar evolution models. These findings are in agreement with, and expand on, earlier work by \citet{Chruslinska:2019dco,Neijssel:2019,Tang:2020,Santoliquido:2021}. Simultaneously comparing the predictions for all three \ac{DCO} flavors can further aid in constraining models, as shown in Figure~\ref{fig:IntrinsicRatesBBHBNSBHNS} for the merger rate densities. 
Another example is that we showed in \S\ref{sec:Detectable-mass-distribution-functions} that the \ac{BHNS} distribution shapes might particularly be a good test bed for stellar evolution models (Figure~\ref{fig:KDE-distributions-BHNS-masses}), whilst the shape of the mass distributions of \ac{BHBH} and \ac{NSNS}  detections are impacted by both stellar evolution and our \SFRD model assumptions (Figure~\ref{fig:KDE-distributions-BHBH-masses}). 
We emphasize that understanding to which uncertainty these different \ac{DCO} observable properties are sensitive to also aids in understanding where to best spend computational time and work in the simulations.    
We showed in \S\ref{sec:results-distributions-BH-mass} that some population synthesis results are robust under our \Nmodels model variations. 
These features are important when identifying distinguishable characteristics of the isolated binary evolution channel compared to other formation channels. 
Constraining and learning from population synthesis models can be further aided in the future by additional constraints including those from redshift dependent rates and redshift dependent distribution shapes \citep[cf.][]{Briel:2021,Chu:2021,Santoliquido:2021} and additional observational constraints including those from electromagnetic observations of  X-ray binaries \citep[e.g.,][]{Belczynski:2020bigBHpaper,Vinciguerra:2020}, Galactic pulsar binaries \citep[e.g.][]{Kruckow:2018,VignaGomez:2018,Chattopadhyay:2020bns,Chattopadhyay:2021bhns}, short gamma-ray bursts \citep[e.g.][]{Mandhai:2021,Zevin:2020sGRBoffset}, or \ac{GW} detections beyond LVK including Cosmic Explorer and \emph{LISA} \citep[e.g.][]{ShaoLi:2018,Ng:2021,Wagg:2021}.

\subsection{Comparison with GW mergers}
\label{sec:discussion-comparison-with-GW-observations}

\begin{figure*}
    \centering
\includegraphics[width=1\textwidth,trim={0cm 0.cm 0 0},clip]{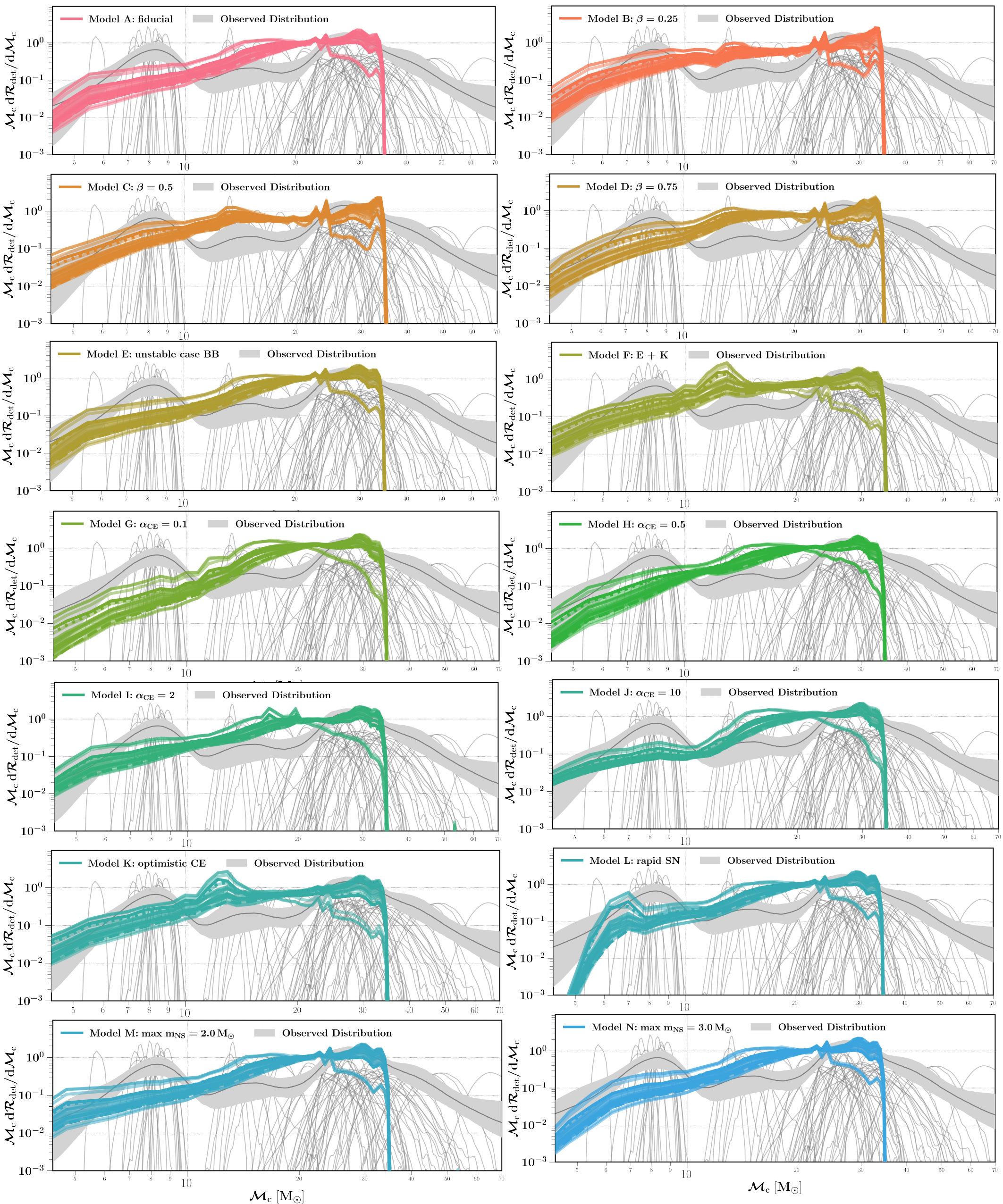} %
\caption{Differential detectable merger rate as a function of chirp mass $ \mchirpf \diff \mathcal{R}_{\rm{det}} / \diff \mchirpf $. Colored distributions show the expected distribution from our simulations. The gray filled distribution shows the $90\%$ interval and median (black line) for the distribution inferred from observations, and in the background the posterior sample distributions for individual BHBH-events are shown \citep[][see Figure 2]{Abbott:2021GWTC3pop}.  All chirp mass distributions are normalized.  \href{https://github.com/FloorBroekgaarden/Double-Compact-Object-Mergers/blob/main/plottingCode/Detectable_Distributions_GWTC-3/COMBINED_part1_loglog.pdf}{\faFileImage} \href{https://github.com/FloorBroekgaarden/Double-Compact-Object-Mergers/blob/main/plottingCode/Detectable_Distributions_GWTC-3/GWTC-3_Fig2.ipynb}{\faBook}
}
    \label{fig:KDE-distributions-BHBH-masses-GWTC-3}
\end{figure*}

\begin{figure*}
    \centering
\includegraphics[width=1\textwidth,trim={0cm 0cm 0 0},clip]{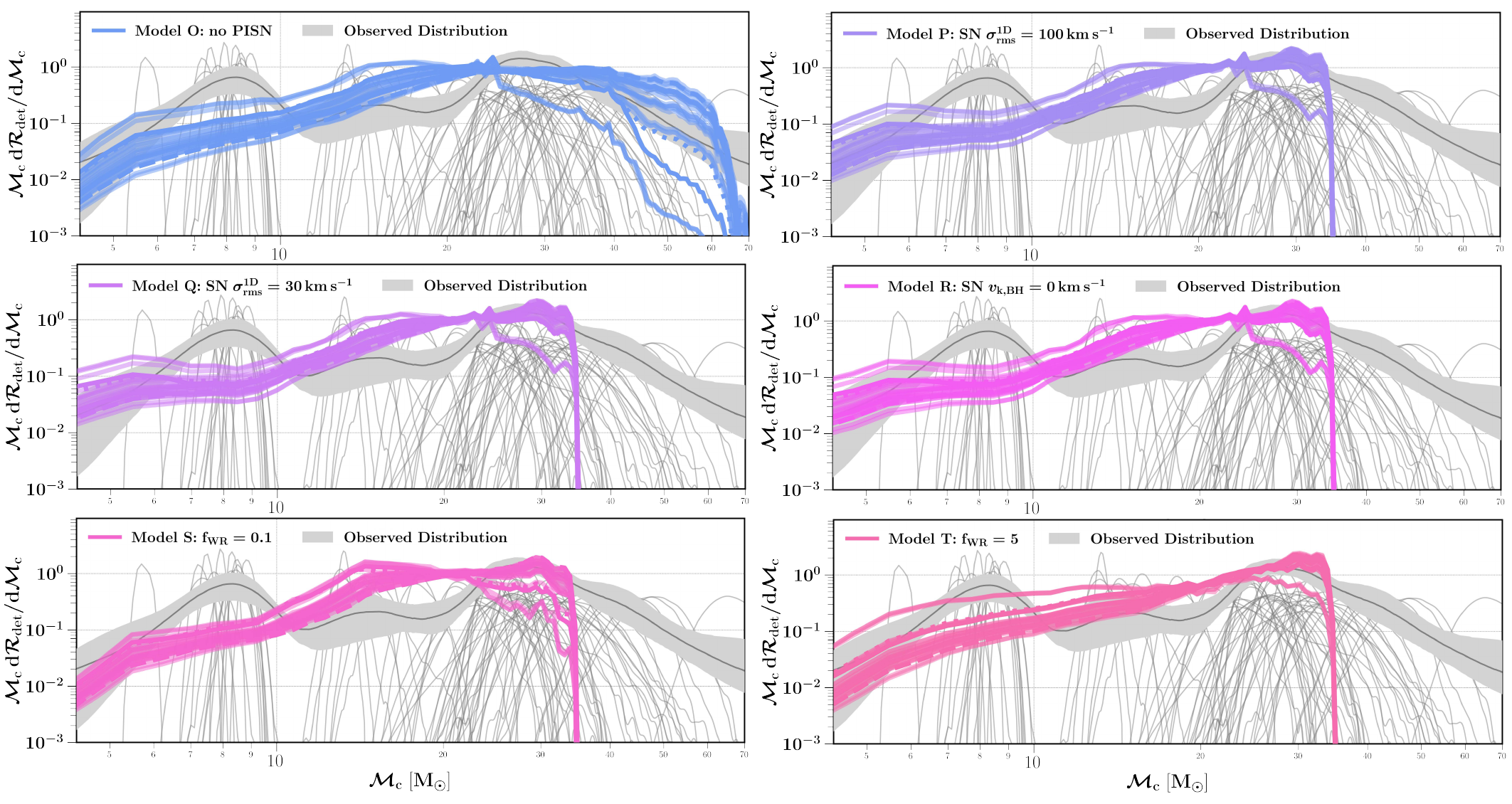} %
\caption*{Continuation of Figure~\ref{fig:KDE-distributions-BHBH-masses-GWTC-3}.  \href{https://github.com/FloorBroekgaarden/Double-Compact-Object-Mergers/blob/main/plottingCode/Detectable_Distributions_GWTC-3/COMBINED_part2_loglog.pdf}{\faFileImage} 
}
    \label{fig:KDE-distributions-BHBH-masses-GWTC-3-part-2}
\end{figure*}

At the time of writing the LVK collaboration published the latest \ac{GW} catalog (GWTC-3; \citealt{Abbott:2021GWTC3}) containing a total of 76 events with a false alarm rate of $\lesssim 1 \yearmin$, consisting of 72 \ac{BHBH} mergers, 2 \ac{NSNS} mergers and 2 \ac{BHNS} events \citet{Abbott:2021GWTC3pop}\footnote{We follow the most likely classification as reported in \citet{Abbott:2021GWTC3pop}. This includes classifying GW190814 as a BHBH event.}. From these detections, the LVK inferred local merger rate densities\footnote{Assuming a merger rate constant in redshift.} of $\rate_{\rm{m}}^{0,\rm{BHBH}}= \RateGWTCminBHBH{-}\RateGWTCmaxBHBH$\GpcminThree\yearmin,  $\rate_{\rm{m}}^{0,\rm{BHNS}} = \RateGWTCminBHNS{-}\RateGWTCmaxBHNS$\GpcminThree\yearmin, and  $\rate_{\rm{m}}^{0,\rm{NSNS}}  =  \RateGWTCminNSNS{-}\RateGWTCmaxNSNS$\GpcminThree\yearmin \citep{Abbott:2021GWTC3pop}. We showed in Figure~\ref{fig:IntrinsicRatesBBHBNSBHNS} that the majority of our \Nmodels model realizations match these inferred local merger rates. However, we also found that a subset of the \SFRD models overestimate the BHBH merger rate, and that most of our models are at the lower end of the NSNS merger rate. If we scale our predicted merger ratios (Figure~\ref{fig:ObservedRatesRatiosBBHBNSBHNS}) to the 72 \ac{BHBH} detections, we find a predicted range of  $0{-}25$ \ac{BHNS}  and $0{-}3$ \ac{NSNS} detections. Although these simulation ranges encompass the \ac{DCO} numbers found in the GWTC-3 catalog, the majority of models seem to particularly underestimate the number of NSNS mergers. This is consistent with findings by other studies on isolated binary evolution populations (e.g., \citealt{GiacobboMapelli:2018,Chruslinska:2018,Santoliquido:2021}), which indicated that matching the observed NSNS rate might require higher values for $\alpha_{\rm{CE}}$ and/or smaller \ac{SN} natal kicks. The NSNS rates could also relatively increase by changes in the mass transfer stability prescriptions ($\zeta$) and envelope binding energy models ($\lambda$), which are not explored in our study \citep[see][and references therein]{Han:2020,Lau:2021}. Future observations and simulations can further constrain this.

Beyond the merger rates, GWTC-3 also provides the inferred mass distributions of the detected BHBH mergers. In Figure~\ref{fig:KDE-distributions-BHBH-masses-GWTC-3} we provide a comparison between the inferred BHBH chirp mass distribution and our set of simulations. We find that the majority of our models predict chirp mass distributions that are largely consistent with the inferred chirp mass distribution (and individual posterior distributions) for $\mchirpf \lesssim 35\Msun$. Figure~\ref{fig:KDE-distributions-BHBH-masses-GWTC-3} shows that particularly the SN remnant mass model variations (model L, M, N and O) significantly impact the lower and higher chirp mass end of the distributions in this comparison. On the other hand, we find that most of model realizations do not match the several detected BHBH mergers with chirp masses $\gtrsim 35$\Msun.  Examples include GW190521 with $\mchirpf = 69.2^{+17.0}_{-10.6}$\Msun, GW190602 with $\mchirpf = 49.1^{+9.1}_{-8.5}$\Msun,  GW190620 with $\mchirpf = 38.3^{+8.3}_{-6.5}$\Msun,  GW190701 with $\mchirpf = 40.3^{+5.4}_{-4.9}$\Msun and GW190706 with $\mchirpf = 42.7^{+10.0}_{-7.0}$\Msun, which all have at least one component with an inferred median BH mass of $\mbhfone \gtrsim 42\Msun$, which is our implemented (pulsational) pair-instability SN BH mass limit. Moreover, even in model O, the model in which we assume that pair-instability \ac{SN} do not occur, we find that $95\% (99\%)$ of the detectable \ac{BHBH} chirp masses are expected to be below $\lesssim 50 (\lesssim 60)\Msun$ (Figure~\ref{fig:KDE-distributions-BHBH-masses}), making it challenging to explain the inferred BH mass in GW190521 and the higher mass range of the chirp mass distribution as shown in Figure~\ref{fig:KDE-distributions-BHBH-masses-GWTC-3}.  The origin of these massive \acp{BH} systems is still under debate.  One possibility is that the location of the pair-instability mass gap could be shifted to higher masses \citep[e.g.][]{2017MNRAS.470.4739S,Farmer:2019, 2020ApJ...902L..36F, 2021MNRAS.501.4514C,Mehta:2021,2021arXiv210307933W}. Another possibility is that BHBH mergers with massive \acp{BH} formed through channels other than the isolated binary evolution channel. For example, \citet[][]{Tanikawa:2021} showed that formation from population III stars can account for the missing high mass BHBH mergers in isolated binary evolution studies like ours that only model population I and II stars. Other formation channels have also been suggested for these massive BHBH mergers including  hierarchical dynamical mergers \citep[e.g.][]{Rodriguez:2015,2016PhRvD..93h4029R, 2020arXiv201006161A}, stellar mergers \citep[e.g.][]{2019MNRAS.485..889S, 2020MNRAS.497.1043D,2020ApJ...903...45K}, triples \citep[e.g.][]{2021ApJ...907L..19V} or mergers in AGN disks \citep[e.g.][]{2020ApJ...903..133S}. For a more detailed discussion see, for example,  \citet{2020ApJ...900L..13A} and \citet{2020arXiv201105332K} and references therein. Future studies should explore a full comparison with LVK data to further constrain models.

\section{Conclusions}
\label{sec:conclusions}

In this study we simulated the rates and properties of \ac{BHBH}, \ac{BHNS} and \ac{NSNS} mergers detectable with existing GW detectors. We simultaneously examined the impact of two key modelling uncertainties in population synthesis studies: uncertainties arising from massive binary star evolution, and from the metallicity-dependent star formation history, \SFRD. We accomplish this by simulating populations of binaries over a grid of 53 birth metallicity values and taking into account the \SFRD and GW detection probability.  The resulting suite of \Nmodels model realizations (\NmodelsBPS binary stellar evolution variations $\times$ \NmodelsMSSFR \SFRD variations) is the largest of its kind, and made publicly available. Our main findings within the context of the variations we considered are summarized below.

\begin{itemize}
    \item \textbf{Merging \ac{DCO} formation yields:} We find for all our stellar evolution variations that the merging BHBH (BHNS) formation yield is typically a rapidly decreasing function of metallicity at   $\Zi \gtrsim \Zsun / 10$ ($\gtrsim \Zsun /2$) as a result of line-driven  Wolf-Rayet like winds. 
    The NSNS yield, on the other hand, is relatively independent of metallicity. 
    We find that the formation yield of BHBH mergers for $\Zi \lesssim \Zsun/10$ is remarkably constant over massive binary-star models and \Zi. 
    The formation yields of BHNS and NSNS mergers are impacted over the full metallicity range by our massive binary-star assumptions  (see Figure~\ref{fig:FormationRateDCO-per-metallicity}).\\

    \item \textbf{Merging \ac{DCO} rates:} We find that the  calculated intrinsic and detectable merger rate densities (\RateIntrinsicZero and \RateObserved, respectively) can be impacted by factors $\approx 210\times$--$1300\times$ due to combined uncertainties in stellar evolution and \SFRD. The \Nmodels model variations lead to estimated merger rate densities in the ranges  
    $\rate^{0,\rm{BHBH}}_{\rm{m}} = \RateIntrinsicAzeroBHBHmin$--$ \RateIntrinsicAzeroBHBHmax \GpcminThree\yearmin$ and
    $\rate^{\rm{BHBH}}_{\rm{det}} = \RateObservedAzeroBHBHmin$--$ \RateObservedAzeroBHBHmax \yearmin$ for \acp{BHBH},      $\rate^{0,\rm{BHNS}}_{\rm{m}} = \RateIntrinsicAzeroBHNSmin$--$\RateIntrinsicAzeroBHNSmax\GpcminThree\yearmin$ and   $\rate^{\rm{BHNS}}_{\rm{det}} = \RateObservedAzeroBHNSmin$--$ \RateObservedAzeroBHNSmax \yearmin$ for \acp{BHNS}, and   $\rate^{0,\rm{NSNS}}_{\rm{m}} = \RateIntrinsicAzeroNSNSmin$--$\RateIntrinsicAzeroNSNSmax \GpcminThree\yearmin$ and   $\rate^{\rm{NSNS}}_{\rm{det}}  \RateObservedAzeroNSNSmin$--$\RateObservedAzeroNSNSmax \yearmin$ for \acp{NSNS}, for a ground-based GW detector consisting of LIGO-Virgo-KAGRA at design sensitivity.
    In particular, we found that the estimated \ac{BHBH} merger rate densities are relatively sensitive to our explored \SFRD uncertainties and might thereby present a good test bed to constrain \SFRD prescriptions.  On the other hand, we found that the \ac{NSNS} merger rate densities are sensitive to stellar evolution models and therefore might be a good test bed to constrain stellar evolution models. 
    For \acp{BHNS} we find that the calculated rates are significantly impacted by both uncertainties  (see Figures~\ref{fig:IntrinsicRatesBBHBNSBHNS} and~\ref{fig:ObservedRatesRatiosBBHBNSBHNS}).\\

    \item \textbf{Merging \ac{DCO} mass distribution shapes:} We show in Figures~\ref{fig:KDE-distributions-BHBH-masses}, ~\ref{fig:KDE-distributions-BHNS-masses} and ~\ref{fig:KDE-distributions-NSNS-masses}  the impact from the massive binary-star and \SFRD variations on the (normalized) shape of the detectable \ac{DCO} mass distributions (chirp mass, individual component masses and mass ratios). 
    We find that the shape of the \ac{BHNS} mass distributions are dominated by variations in binary stellar evolution within our model explorations. 
    For \ac{BHBH} and NSNS mergers we find that the mass distribution shapes are impacted by both variations in stellar evolution and \SFRD.\\

    \item \textbf{Merging \ac{DCO} \tdelay and \Zi distribution shapes:} We show in  Figures~\ref{fig:KDE-distributions-BHBH-delay-times}, ~\ref{fig:KDE-distributions-BHNS-delay-times} and~\ref{fig:KDE-distributions-NSNS-delay-times} the impact from binary stellar evolution and \SFRD variations on the delay time and birth metallicity distributions calculated in our models. 
    We find that the \SFRD has a significant impact for BHBH, BHNS and NSNS on the shape of the delay time and metallicity distributions of detectable mergers.  
    Several stellar evolution models, including those affecting stellar winds, can also significantly impact the delay time and \Zi distribution shapes.\\

    \item \textbf{Consistent features among our \Nmodels model variations:} We find several features in the \ac{DCO} mass distributions that are consistent among all our \Nmodels model variations. 
    First, we find that at least $95\%(99\%)$ of the detectable \ac{BHBH} mergers have mass ratios ${\qf\lesssim 4 (\qf \lesssim 6)}$ in all \Nmodels model variations (fifth column Figure~\ref{fig:KDE-distributions-BHBH-masses}). Second, we find that more than $95\%$ of \ac{BHBH} mergers are always expected to contain a \ac{BH} with a mass $\gtrsim 8\Msun$ (third column Figure~\ref{fig:KDE-distributions-BHBH-masses}). 
    Third, we find that less than $5\%$  of the detectable merging \ac{BHNS} systems have \ac{BH} masses ${\mbhf \gtrsim 18\Msun}$ (third column Figure~\ref{fig:KDE-distributions-BHNS-masses}). Fourth we find that  \acp{NS} in NSNS mergers are expected to have lower masses on average compared to \acp{NS} in \ac{BHNS} mergers (Figures~\ref{fig:KDE-distributions-BHNS-masses} and ~\ref{fig:KDE-distributions-NSNS-masses}).
    We discuss how these findings are marginally consistent with GW observations from GWTC-3 in \S\ref{sec:results-distributions-BH-mass} and \S\ref{sec:discussion-comparison-with-GW-observations}.
\end{itemize}

Overall, our results highlight the importance of considering the uncertainty in both the stellar evolution and metallicity-dependent star formation history when exploring population synthesis simulations of BHNS, BHBH and NSNS mergers and when trying to infer model properties from GW data.

\section*{Acknowledgements}

The authors thank everyone in the COMPAS collaboration and Berger Time-Domain Group for help. In addition, the authors thank the Harvard FAS research computing group for technical support on the simulations and high performance computing part of the research. The authors also thank Shanika Galaudage and Victoria DiTomasso for their help with this paper. FSB thanks Christopher Brown and Katie Callam for organizing the `writing oasis'.   Lastly, FSB wants to acknowledge the  amount of serendipity and privilege that was involved to end up pursuing this astronomy research. The Berger Time-Domain Group is supported in part by NSF and NASA grants. Some of the authors are supported by the Australian Research Council Centre of Excellence for Gravitational Wave Discovery (OzGrav), through project number CE170100004. FSB is supported in part by the Prins Bernard Cultuurfonds studiebeurs 2021. IM is a recipient of the Australian Research Council Future Fellowship FT190100574. A.V-G. acknowledges funding support by the Danish National Research Foundation (DNRF132). SJ, LvS and SdM acknowledge funding from the Netherlands Organisation for Scientific Research (NWO), as part of the Vidi research program BinWaves (project number 639.042.728) and the European Union’s Horizon 2020 research and innovation program from the European Research Council (ERC, Grant agreement No. 715063). LvS, TW and SdM acknowledge support by the National Science Foundation under Grant No. (NSF 2009131). This research has made use of NASA’s Astrophysics Data System Bibliographic Services. This research has made use of data, software and/or web tools obtained from the Gravitational Wave Open Science Center (\url{https://www.gw-openscience.org/}), a service of LIGO Laboratory, the LIGO Scientific Collaboration and the Virgo Collaboration. LIGO Laboratory and Advanced LIGO \citep{2015CQGra..32g4001L} are funded by the United States National Science Foundation (NSF) as well as the Science and Technology Facilities Council (STFC) of the United Kingdom, the Max-Planck-Society (MPS), and the State of Niedersachsen/Germany for support of the construction of Advanced LIGO and construction and operation of the GEO600 detector. Additional support for Advanced LIGO was provided by the Australian Research Council. Virgo \citep{2015CQGra..32b4001A} is funded, through the European Gravitational Observatory (EGO), by the French Centre National de Recherche Scientifique (CNRS), the Italian Istituto Nazionale di Fisica Nucleare (INFN) and the Dutch Nikhef, with contributions by institutions from Belgium, Germany, Greece, Hungary, Ireland, Japan, Monaco, Poland, Portugal, Spain \citep{ligo_scientific_collaboration_and_virgo_2021_5655785}.

\section*{Data availability}
\label{sec:data-availability}
All data used in this work is publicly available on Zenodo at \citet[][BHBH]{Broekgaarden:2021-zenodo-BHBH} \citet[][BHNS]{Broekgaarden:2021-zenodo-BHNS} and  \citet[][NSNS]{Broekgaarden:2021-zenodo-NSNS}. All code to reproduce the results and figures in this paper (and additional figures) are publicly available on Github at \url{https://github.com/FloorBroekgaarden/Double-Compact-Object-Mergers} \href{https://github.com/FloorBroekgaarden/Double-Compact-Object-Mergers}{\faGithub}.

\section*{Software} 
Simulations in this paper made use of the COMPAS rapid binary population synthesis code, which is freely available at \url{http://github.com/TeamCOMPAS/COMPAS} \citep{COMPAS:2021methodsPaper} including work based on  \citep{stevenson2017formation, 2018MNRAS.477.4685B, VignaGomez:2018,Broekgaarden:2019}. 
The simulations performed in this work were simulated with a COMPAS version that predates the publicly available code. Our version of the code is most similar to version 02.13.01 of the publicly available COMPAS code. 
Requests for the original code can be made to the lead author. 
The authors used {\sc{STROOPWAFEL}} from \citep{Broekgaarden:2019}, publicly available at \url{https://github.com/FloorBroekgaarden/STROOPWAFEL}\footnote{For the latest pip installable version of STROOPWAFEL please contact the corresponding author.}.

The authors made use of \textsc{Python} from the  Python Software Foundation. Python Language Reference, version 3.6. Available at \url{http://www.python.org} \citep{CS-R9526}. In addition the following Python packages were used: \textsc{matplotlib} \citep{2007CSE.....9...90H},  \textsc{NumPy} \citep{2020NumPy-Array}, \textsc{SciPy} \citep{2020SciPy-NMeth}, \texttt{ipython$/$jupyter} \citep{2007CSE.....9c..21P, kluyver2016jupyter}, \textsc{pandas} \citep{mckinney-proc-scipy-2010}, \textsc{Seaborn} \citep{waskom2020seaborn}, \textsc{Astropy} \citep{2018AJ....156..123A}  and   \href{https://docs.h5py.org/en/stable/}{\textsc{hdf5}} \citep{collette_python_hdf5_2014}. 

Figure~\ref{fig:KDE-distributions-BHBH-masses-GWTC-3} makes use of the \textsc{make$\_$plots$\_$Vamana.py} code by Vaibhav Tiwari on behalf of the LIGO Scientific Collaboration, Virgo Collaboration and KAGRA Collaboration provided under the Creative Commons Attribution 4.0 licence. We obtained their code from Zenodo \url{doi:10.5281/zenodo.5655785}, which makes use of the kernel density estimator as provided in \url{https://dcc.ligo.org/LIGO-T2100447/public} \citep{Sadiq:2021kde}. 
The simulations were performed on the super computers from the Harvard FAS research computing group.



\bibliographystyle{mnras}
\bibliography{DCO} 



\appendix

\section{Additional tables and figures}
In this appendix we show a Table and three additional figures that are mentioned throughout the main text.

First, Table~\ref{tab:MSSFR-variations-labels} summarizes the \NmodelsMSSFR \SFRD models that are explored in this study. More details about the assumptions in each model can be found in \citet[][]{Broekgaarden:2021} and \citet{Neijssel:2019}.

\begin{table*}
\resizebox{\textwidth}{!}{%
\centering
\begin{tabular}{llll}
\hline
\hline
{xyz} index &  \ac{SFRD} [x]                     & \ac{GSMF} [y]       & \ac{MZR} [z]                        \\ \hline \hline
000  				& \multicolumn{3}{c}{preferred phenomenological  model from    \citet{Neijssel:2019} }          \\
\hline
1              			       	& {\citet{2014ARA&A..52..415M}}  &  \citet{2004MNRAS.355..764P} & \citet{2006ApJ...638L..63L}   \\
2              					&  \citet{2004ApJ...613..200S}							& \citet{2015MNRAS.450.4486F} single Schechter   & \citet{2006ApJ...638L..63L} $+$ offset    \\
3              			      	& \citet{2017ApJ...840...39M}     		& \citet{2015MNRAS.450.4486F} double Schechter          &  \citet{2016MNRAS.456.2140M}             \\ \hline
\end{tabular}%
}
\caption{List of the assumptions for the metallicity-specifc star formation rate, \SFRD, models  that we explore in this study. 27 \SFRD models are obtained by combining a star formation rate density (SFRD) with a galaxy stellar mass function (GSMF) and mass-metallicity relation (MZR). 
See  \S\ref{sec:method-metallicity-specific-star-formation-rate-density-prescription} and paper I for more details. 
The models are named in the convention $\rm{xyz}$, where  $\rm{x, y, z} \in [1,2,3]$ are the index numbers for the models used for the  \ac{SFRD}, \ac{GSMF} and \ac{MZR}, respectively.  
For example, the combination of using the {\citet{2014ARA&A..52..415M}} \ac{SFRD}  with the  \citet{2004MNRAS.355..764P} \ac{GSMF} and the  \citet{2016MNRAS.456.2140M}  \ac{MZR}  is labeled ${113}$. 
The preferred phenomenological model from  \citet{Neijssel:2019} has the label ${000}$ and is not a specific combination but a parameterized model that is built to be flexible and is fitted to match the observed \ac{BHBH} merger rate and chirp mass distribution from the first two runs of LIGO and Virgo. }
\label{tab:MSSFR-variations-labels}
\end{table*}

\begin{figure*}
    \centering
\includegraphics[width=1.0\textwidth]{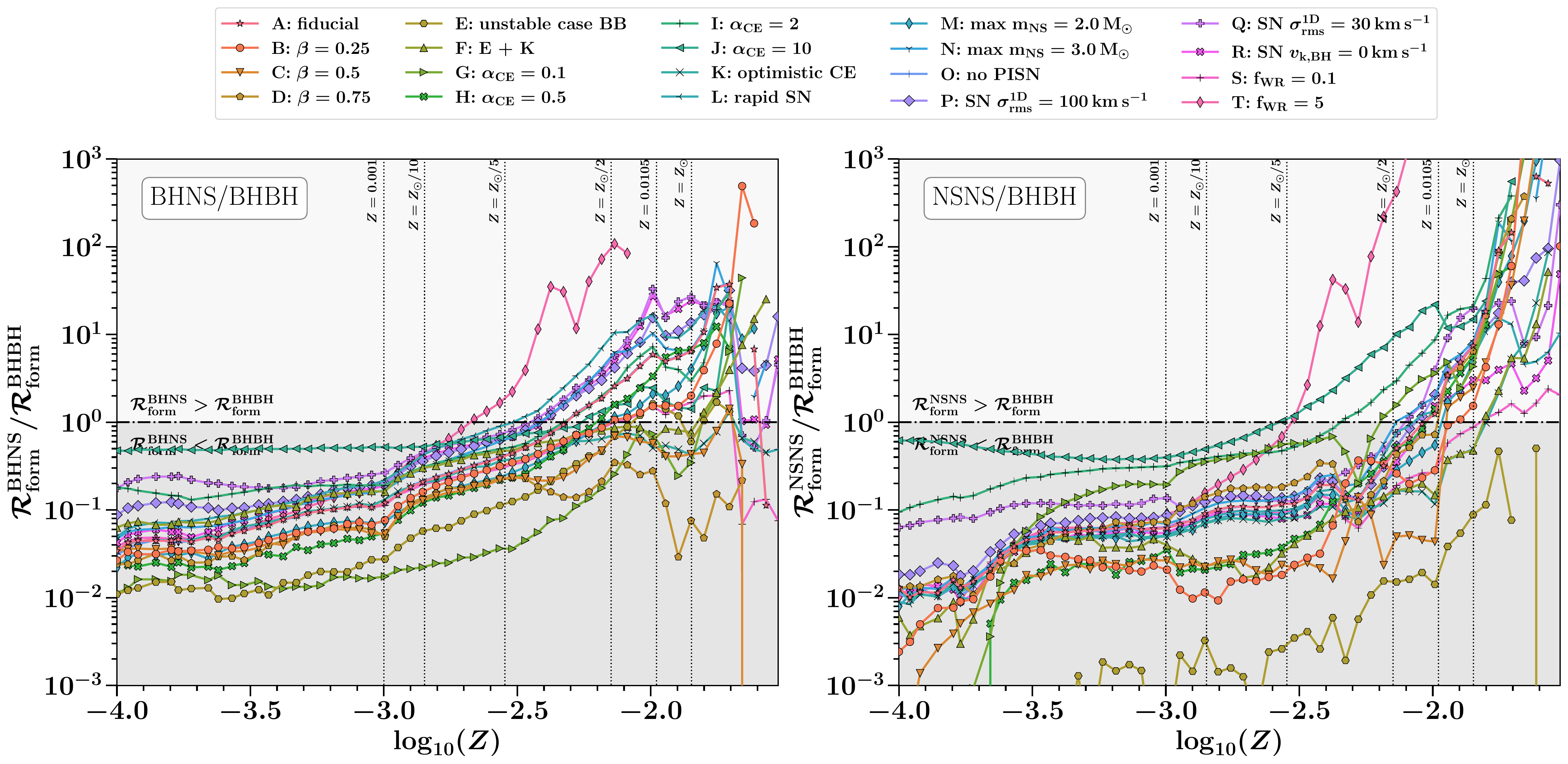}
\caption{The ratio of formation yields of merging double compact objects per solar mass of star formation ($\diff N_{\rm{form}} / \diff \MSFR$) as a function of birth metallicity (\Zi). 
The left panel shows the ratio of BHNS to BHBH formation yields and the right panel shows the ratio of NSNS to BHBH formation yields.  Each color and marker type corresponds to one of the \NmodelsBPS binary population synthesis models explored in this study (Table~\ref{tab:variations-BPS}). 
Vertical dotted lines show fixed \Zi values. 
The marker points show the \Zi grid points that we simulated with COMPAS. Scatter points are missing where we divide by zero. \href{https://github.com/FloorBroekgaarden/Double-Compact-Object-Mergers/blob/main/plottingCode/Fig_1_and_Fig_A1/FormationRatioRateAllModels3panels.pdf}{\faFileImage} \href{https://github.com/FloorBroekgaarden/Double-Compact-Object-Mergers/blob/main/plottingCode/Fig_1_and_Fig_A1/make_Fig_1_and_Fig_A1.ipynb}{\faBook}} 
\label{fig:FormationRateDCO-Ratios-per-metallicity}
\end{figure*}

Second, in Figure~\ref{fig:FormationRateDCO-Ratios-per-metallicity} we show similar to Figure~\ref{fig:FormationRateDCO-per-metallicity} the formation yield of \ac{DCO} systems that merge in a Hubble time as a function of the birth metallicity. However, in this figure we show the relative formation yields between \ac{BHNS} to \ac{BHBH} (left panel) and between \ac{NSNS} and \ac{BHBH} (right panel). It can be seen that in most models the formation yield of \ac{BHNS} mergers exceeds that of \ac{BHBH} around birth metallicities $\Zi \gtrsim \Zi/2$. The yield of \ac{NSNS} mergers surpasses the \ac{BHBH} yield for most models around $\Zi / \Zsun$.

\begin{figure*}
    \centering
        \subfloat[][The $\rm{xyz}=312$ \SFRD model. \label{fig:SFRD-000}]{{\includegraphics[width=1\columnwidth]{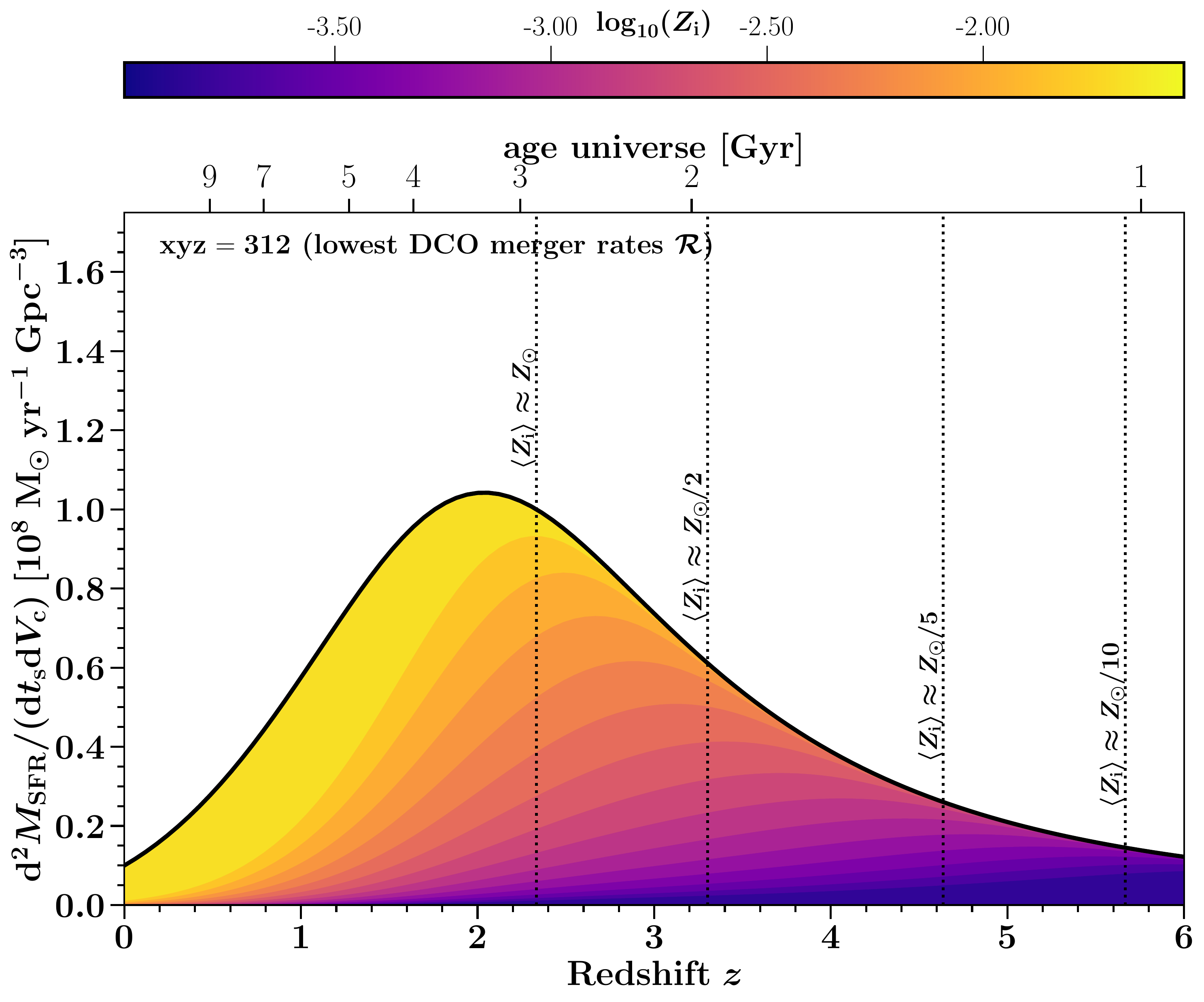} }}%
        \qquad
        \subfloat[][The $\rm{xyz}=231$ \SFRD model. \label{fig:SFRD-231}]{{\includegraphics[width=1\columnwidth]{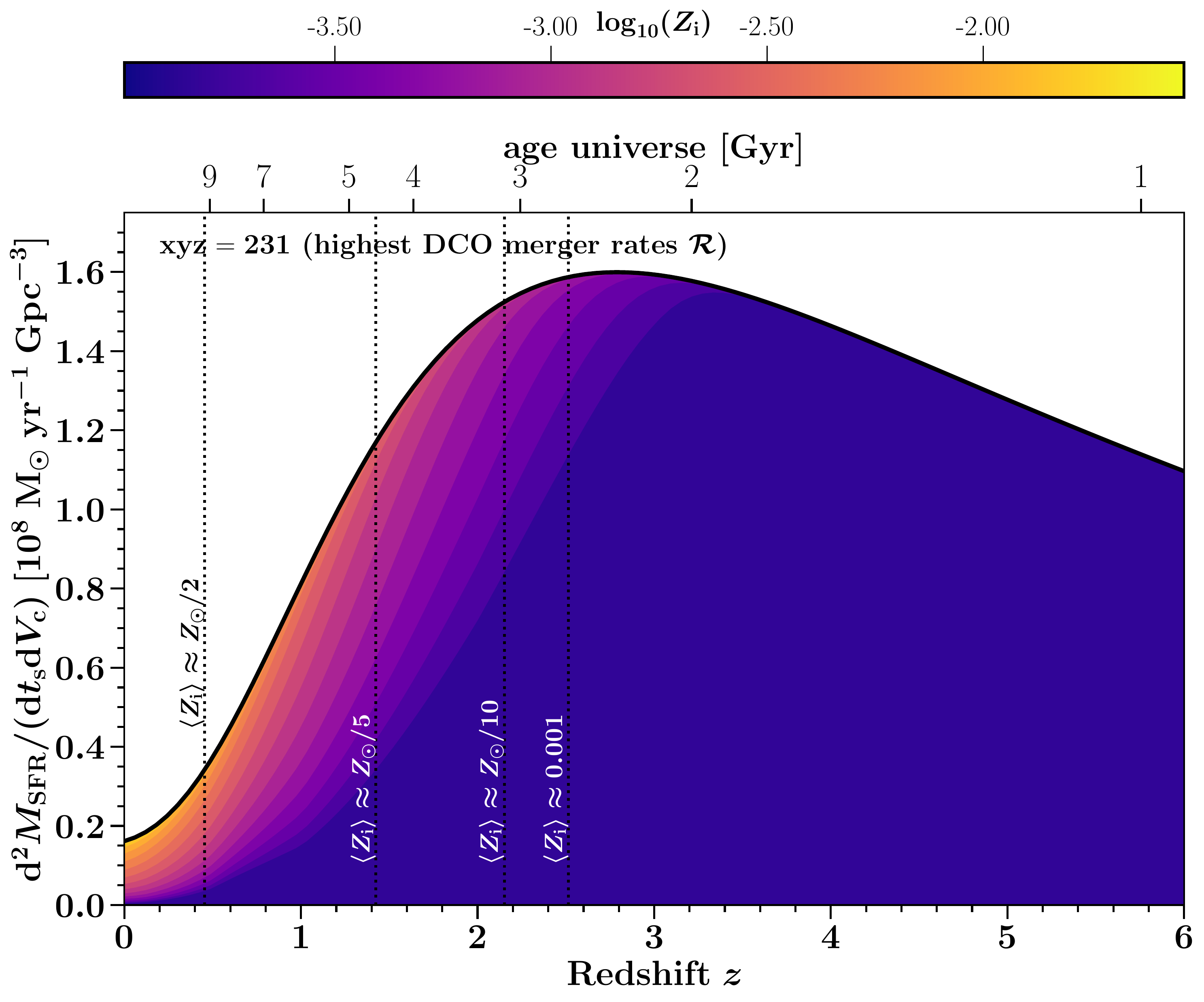} }}%
      \qquad
  \caption{The \ac{SFRD} as a function of redshift and metallicity for two \SFRD models from Table~\ref{tab:MSSFR-variations-labels} used in this study. The left panel shows the  $\rm{xyz}=312$    \SFRD model whilst the right panel shows the  $\rm{xyz}=231$ \SFRD model, these correspond to the \SFRD models with the highest and lowest average \Zi, respectively. In color we show the build up of the \ac{SFRD} from the contribution by different \Zi, where  for visual purposes we divided the \ac{SFRD} into a few  metallicity bins.  \href{https://github.com/FloorBroekgaarden/Double-Compact-Object-Mergers/blob/main/plottingCode/Fig_A2/SFRDandZi_xyz312_.pdf}{\faFileImage} \href{https://github.com/FloorBroekgaarden/Double-Compact-Object-Mergers/blob/main/plottingCode/Fig_A2/SFRDandZi_xyz231_.pdf}{\faFileImage}  \href{https://github.com/FloorBroekgaarden/Double-Compact-Object-Mergers/blob/main/plottingCode/Fig_A2/make_Fig_A2.ipynb}{\faBook}}
  \label{fig:2panels_examples_SFRD-Z}
\end{figure*}

Third, in Figure~\ref{fig:2panels_examples_SFRD-Z} we visualize two of our \SFRD model variations, $\rm{xyz} = 312$ and $\rm{xyz} = 231$. The $\rm{xyz} = 312$ ($\rm{xyz} = 231$ ) model is the \SFRD distribution that correspond to  one of the highest (lowest) average \Zi and results in one of the lowest (highest) \ac{BHBH} merger rates in Figure~\ref{fig:IntrinsicRatesBBHBNSBHNS}. 

\begin{figure*}
    \centering
\includegraphics[width=0.8\textwidth]{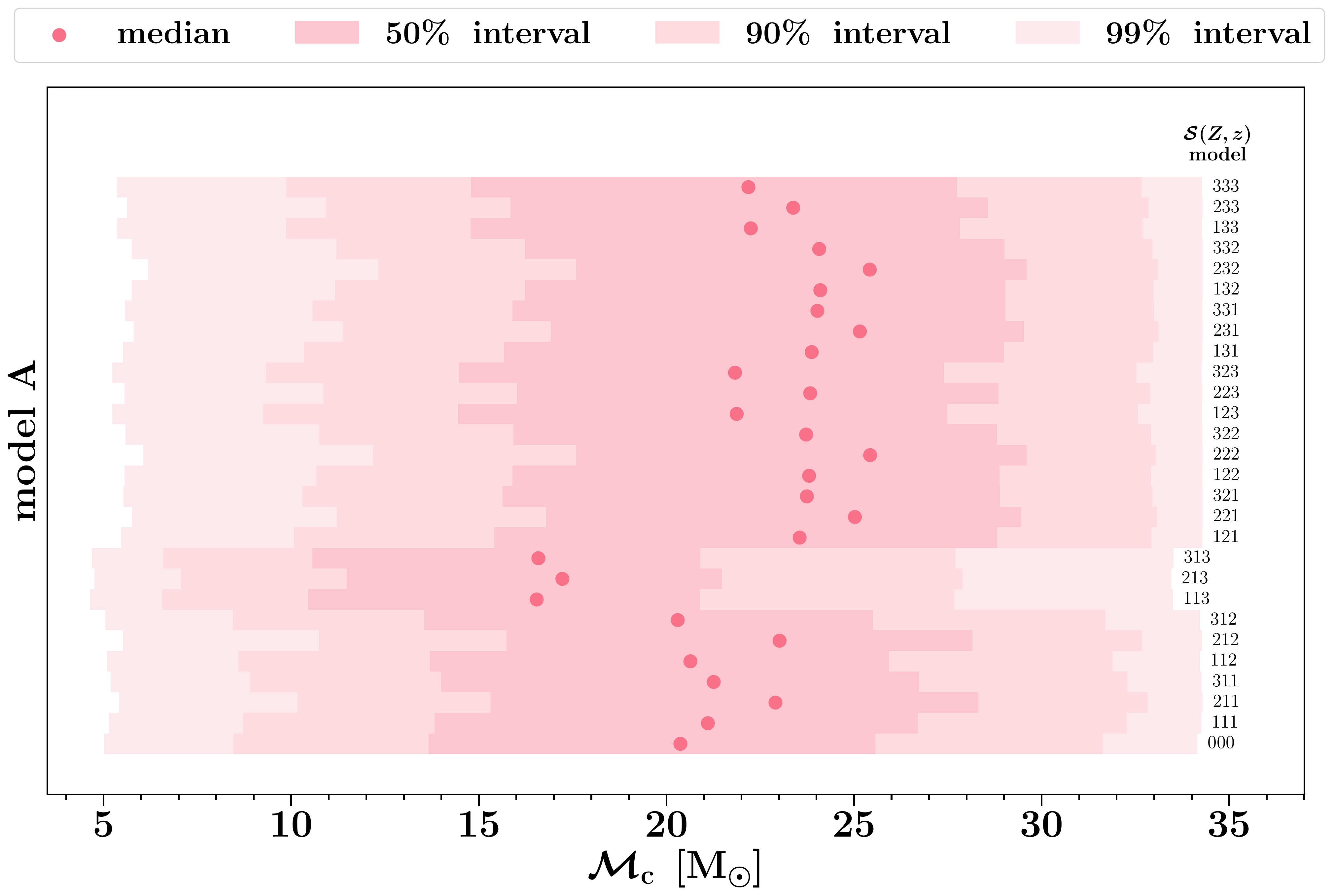}
\caption{ Zoom in on the chirp mass percentile panel for model A for detectable BHBH mergers at design sensitivity from Figure~\ref{fig:KDE-distributions-BHBH-masses}.  Each of the \NmodelsMSSFR  individual horizontal bars  visualizes the median (scatter points), $50$ , $90$  and $99$  (three shades) distribution intervals. On the right of the horizontal bar we show the \SFRD model name as given in Table~\ref{tab:MSSFR-variations-labels}.  \href{https://github.com/FloorBroekgaarden/Double-Compact-Object-Mergers/blob/main/plottingCode/Fig_4_and_Fig_5_and_Fig_6/PercentilePlot_ZOOM.pdf}{\faFileImage}  \href{https://github.com/FloorBroekgaarden/Double-Compact-Object-Mergers/blob/main/plottingCode/Fig_4_and_Fig_5_and_Fig_6/make_figure_A3.ipynb}{\faBook}  } 
    \label{fig:chirp-mass-percentile-zoom-in}
\end{figure*}

Last, Figure~\ref{fig:chirp-mass-percentile-zoom-in} shows a zoom in on one of the percentile distribution panels from Figure~\ref{fig:KDE-distributions-BHBH-masses} (the top row, second column panel). We highlight in this figure the labels of the \SFRD models that correspond to the order of the horizontal percentile bars. 

%


\bsp	
\label{lastpage}
\end{document}